\documentclass[twocolumn]{aastex631}

\usepackage{graphicx}	
\usepackage{amsmath}	
\usepackage[version=4]{mhchem}
\usepackage{gensymb}
\usepackage{chemfig}

\received{YY}
\revised{YY}
\accepted{YY}

\submitjournal{ApJ}

\shorttitle{longwave scattering approximations for exoplanets}
\shortauthors{Elspeth K. H. Lee}
\graphicspath{{./}{plots/}}

\begin{document}

\title{Testing approximate infrared scattering radiative-transfer methods for hot Jupiter atmospheres}

\correspondingauthor{Elspeth K.H. Lee}
\email{elspeth.lee@unibe.ch}

\author[0000-0002-3052-7116]{Elspeth K.H. Lee}
\affiliation{Center for Space and Habitability, University of Bern, Gesellschaftsstrasse 6, CH-3012 Bern, Switzerland}

\begin{abstract}
The calculation of internal atmospheric (longwave) fluxes is a key component of any model of exoplanet atmospheres that requires radiative-transfer (RT) calculations.
For atmospheres containing a strong scattering component such as cloud particles, most 1D multiple-scattering RT methods typically involve numerically expensive matrix inversions.
This computational bottleneck is exacerbated when multitudes of RT calculations are required, such as in general circulation models (GCMs) and retrieval methods.
In an effort to increase the speed of RT calculations without sacrificing too much accuracy, we investigate the applicability of approximate longwave scattering methods developed for the Earth science community to hot Jupiter atmospheres.
We test the absorption approximation (AA) and variational iteration method (VIM) applied to typical cloudy hot Jupiter scenarios, using 64 stream DISORT calculations as reference solutions.
We find the four-stream VIM variant is a highly promising method to explore using for hot Jupiter GCM and retrieval modelling, showing excellent speed characteristics, with typical errors $\sim$1\% for outgoing fluxes and within $\sim$50\%, but with larger errors in the deep cloud layer test case, for vertical heating rates. 
Other methods explored in this study were found to typically produce similar error characteristics in vertical heating rates.
\end{abstract}


\section{Introduction} 
\label{sec:intro}

The calculation of thermal internal atmospheric fluxes, traditionally called `longwave' radiation, is a key part of investigating the emission properties of exoplanet atmospheres.
This type of calculation is critical to modelling and fitting secondary eclipse (dayside) spectral data of tidally locked exoplanets \citep[e.g.][]{Crouzet2014,Kreidberg2014,Coulombe2023} as well as full orbit thermal spectral phase curves of exoplanets \citep[e.g.][]{Stevenson2014,Arcangeli2019,Mikal-Evans2022}.

With the advent of JWST, exoplaneteers now have access to an expansive wavelength range, unparalleled sensitivity and continuous staring time for these types of observations than previous keystone observatories.
However, this comes with an increase in the challenge of appropriately modelling this data for both forward and retrieval efforts.
The wavelength dependent opacity and scattering properties of potential clouds and hazes in these atmospheres can have a large impact on an objects spectra and interpretation of observational data.

Several recent studies have shown the importance of incorporating longwave scattering when modelling cloud properties.
\citet{Kitzmann2013} compare a 24-stream DISORT calculation with a two-stream method to investigate the longwave greenhouse effect of CO$_{2}$ clouds on exoplanet atmospheres. 
They find large deviations between the DISORT and two-stream methods alters the strength of the longwave greenhouse effect, impacting the climate of potential worlds containing CO$_{2}$ clouds.
\citet{Taylor2021} provide analytical expectations for the impact of a scattering component on the outgoing flux of hot Jupiter atmospheres.
They show that degeneracies and biases in retrieval results can occur when the longwave scattering is not taken into account and propose a method to retrieve optical constants of the scattering component.
\citet{Taylor2023} explore the possible presence of infrared scattering clouds on the dayside of HD 209458b and WASP-43b through retrieval modelling, finding that HD 209458b shows no evidence of a thermal scattering component, while the evidence for WASP-43b is dependent on the specific data reduction used.
 
Typically, 1D column approaches such as radiative-convective-equilibrium (RCE) modelling are numerically expedient enough to include more sophisticated RT frameworks \citep[e.g.][]{Drummond2016,Molliere2017,Gandhi2019,Malik2019}.
However, 1D retrieval modelling of exoplanet emission spectra data requires many (sometimes millions) of RT iterations and computational efficiency is an important consideration when developing retrieval models \citep[e.g. for emision spectra;][]{Line2014,Waldmann2015,Gandhi2018,Kitzmann2020,Cubillos2022}.
With JWST data, the increased spectral resolution as well as spectral wavelength range has substantially increased the RT model burden on retrieval efforts.
With the observation of data rich phase curves of exoplanets, consistent 2D retrievals from thermal phase curve data has also been approached \citep[e.g.][]{Irwin2020,Feng2020,Changeat2021,Dobbs-Dixon2022}, further increasing the number of RT calculations required to be performed.

3D General Circulation Models (GCMs) are also a class of models that require fast RT schemes. 
Typically, a 1D RT model is applied for every vertical column of the GCM, which for usual hot Jupiter GCM grid resolutions is thousands of columns (e.g. 6144 for the C32 cubed-sphere resolution commonly used by SPARC/MITgcm \citep{Showman2009}) requiring calculation.
For an integration time of 100 simulated Earth days, the total number of RT calculations can be in the 10s of million range depending on the radiative timestep used.

In this brief study, we investigate the use of longwave scattering approximation methods in use in the Earth atmospheric sciences and apply them to typical hot Jupiter conditions.
In Section \ref{sec:RT_eq}, we briefly review the RT equation for longwave radiation propagation. 
Section \ref{sec:AA} provides the approximate solution of RT equation, the absorption approximation (AA) and extended absorption approximation (EAA). 
Section \ref{sec:VIM} shows the variational iteration method (VIM).
In Section \ref{sec:tests}, we test each method for a cloudy HD 189733b scenario, comparing the outgoing longwave radiation (OLR) flux and vertical heating rates to 64 stream DISORT reference calculations.
Section \ref{sec:disc} contains the discussion and Section \ref{sec:conc} the conclusions of our study.

\section{Radiative-Transfer equation}
\label{sec:RT_eq}

At the core of the longwave radiative-transfer equation in plane-parallel geometry is an expression of energy conservation through a medium by thermal and scattering processes.
The radiative-transfer equation for the thermal component in a plane-parallel atmosphere is expressed as \citep[e.g.][]{Li2002}
\begin{multline}
  \mu\frac{\partial I(\tau_{0}, \mu)}{\partial \tau_{0}} = I(\tau_{0}, \mu) - (1 - \omega_{0}) B(\tau_{0}) \\ - \frac{\omega_{0}}{2}\int_{-1}^{1}I(\tau_{0}, \mu')P(\mu, \mu')d\mu',
\end{multline}
where $\mu$ = $\cos\theta$ is the cosine angle of propagation of the beam, $\tau_{0}$ the optical depth (used as the vertical coordinate system), I [erg s$^{-1}$ cm$^{-2}$ sr$^{-1}$] the intensity of the radiation, $\omega_{0}$ the single scattering albedo, B(T) [erg s$^{-1}$ cm$^{-2}$ sr$^{-1}$] the Planck function at temperature T [K] and P($\mu$, $\mu'$) the scattering phase function, which gives the probability of radiation being scattered from the $\mu'$ direction towards the $\mu$ beam direction.

The term with the Planck function contains the contribution of medium's thermal energy to the intensity of the beam. 
This is assumed to be an isotropic source, so does not contain any directional dependent component.
The second term contains the contribution of scattered light from other directions.
This is governed by the properties of the scattering phase function P($\mu$, $\mu'$), which can take many forms, but usually assumed to either be isotropic or equal to the Henyey-Greenstein analytic phase function \citep{Henyey1941}.
This is integrated across all $\mu'$ directions, providing the net contribution of scattered light towards a given $\mu$ beam direction.
In the following sections, we attempt to keep the expressions generally consistent with the \citet{Heng2014} and associated papers notation, adding the concepts for each approximation following \citet{Li2000} and \citet{Zhang2017}.

\section{Absorption Approximation}
\label{sec:AA}

In the absorption approximation (AA), opacity sources are assumed to only contribute their absorption opacity to the atmospheric transmission, neglecting their scattering component to the total extinction opacity.
In this approach, the scattering source function is ignored and incorporated into an alteration to the extinction opacity of the atmosphere, this reduces the RT equation to \citep[e.g.][]{Zhang2017}
\begin{equation}
\label{eq:AA}
    \mu\frac{dI(\tau_{0}, \mu)}{d\tau_{0}} = \epsilon \left[I(\tau_{0}, \mu) - B(\tau_{0})\right],
\end{equation}
where $\epsilon$ = 1 - $\omega_{0}$ is co-albedo.

To take into account non-isothermal layers an exponential functional form for B($\tau_{0}$) is used \citep[e.g.][]{Fu1993, Zhang2017}
\begin{equation}
\label{eq:expB}
    B(\tau_{0}) \approx B_{1}\exp({-\beta\tau_{0}}),
\end{equation}
where $\beta$ = -[$\ln(B_{2}/B_{1})]/\tau$, $\tau$ = $\tau_{2}$ - $\tau_{1}$ is the optical depth of the layer. 
Following \citet{Heng2014}, here the index 1 denotes the upper level (located at lower optical depth) and 2 the lower level (located at higher optical depth).
A commonly used functional form is the so-called linear in $\tau$ approximation given by \citep[e.g.][]{Toon1989}
\begin{equation}
    B(\tau_{0}) \approx B_{1} + \frac{B_{2} - B_{1}}{\tau}.
\end{equation}
\citet{Li2002} discuss the differences between the exponential and linear forms.

We define the regular transmission function, $\mathcal{T}$, of an atmospheric layer as 
\begin{equation}
    \mathcal{T}(\mu) = \exp(-\tau/\mu),
\end{equation}
and the absorption only transmission function, $\mathcal{T}_{\rm a}$, of an atmospheric layer as 
\begin{equation}
    \mathcal{T}_{\rm a}(\mu) = \exp(-\epsilon\tau/\mu).
\end{equation}

The solution of Eq. \ref{eq:AA}, assuming the exponential Planck function form (Eq. \ref{eq:expB}), is \citep[e.g.][]{Zhang2017}
\begin{equation}
    I_{\uparrow,1}(\mu_{i}) = I_{\uparrow,2}(\mu_{i})\mathcal{T}_{a} + \frac{\epsilon}{\mu\beta + \epsilon}\left[B_{1}- B_{2}\mathcal{T}_{a}\right],
\end{equation}
\begin{equation}
    I_{\downarrow,2}(\mu_{-i}) = I_{\downarrow,1}(\mu_{-i})\mathcal{T}_{a} + \frac{\epsilon}{\mu\beta - \epsilon}\left[B_{1}\mathcal{T}_{a} - B_{2}\right],
\end{equation}
for the upward and downward directions respectively.
The upper boundary condition at layer index 0 is typically
\begin{equation}
     I_{\downarrow, 0} = 0,
\end{equation}
unless some other thermal source is required to be taken into account above the computational boundary.
For a gas giant atmosphere with given internal temperature, T$_{\rm int}$ [K], the lower boundary condition is given by
\begin{equation}
\label{eq:Tint}
    I_{\uparrow, n} = I_{\downarrow, n} + B(T_{\rm int}),
\end{equation}
where n is the index of the lowest boundary level.
This condition ensures energy conservation across the boundary and the net flux is equal to the internal thermal energy.
Integration proceeds by first calculating the downward intensity from the upper boundary to the lower, then using the downward intensity at the lowest level through Eq. \ref{eq:Tint} to the perform the return upward intensity sweep.

The AA method is simple to implement as there is no coupling between the upward and downward direction from a scattering component. 
Overall, AA effectively replaces the extinction opacity used in the regular transmission function with the absorption opacity, meaning any method that doesn't compute a scattering source function can use AA in a simple manner through a change in the transmission function.

The upward and downward flux, F$_{\uparrow, \downarrow}$ in each layer is calculated using Gaussian quadrature through
\begin{equation}
    F_{\uparrow} = 2\pi\int_{0}^{1} I_{\uparrow}d\mu = 2\pi\sum_{n}w_{n}\mu_{n}I_{\uparrow}(\mu_{n}),
\end{equation}
\begin{equation}
    F_{\downarrow} = 2\pi\int_{-1}^{0}I_{\downarrow}d\mu = 2\pi\sum_{n}w_{n}\mu_{n}I_{\downarrow}(\mu_{-n}),
\end{equation}
respectively.
For the two-stream variant (2AA) a single Gaussian quadrature value and weight is used, traditionally $\mu_{1}$ = 1/1.66, w$_{1}$ = 1 following \citet{Elsasser1942}.
The four-stream variant (4AA) uses a double Gaussian quadrature scheme with $\mu_{1}$ = 0.21132487, w$_{1}$ = 0.5 and $\mu_{2}$ = 0.78867513, w$_{2}$ = 0.5, which comes at the additional cost of performing four sweeps in the vertical column for the flux calculation.

Lastly, the heating rate $\partial$T/$\partial$t [K s$^{-1}$] for a layer is given by
\begin{equation}
    \frac{\partial T}{\partial t} = \frac{g}{c_{\rm p}} \frac{\partial F}{\partial p},
\end{equation}
where g [cm s$^{-2}$] is the gravitational acceleration, c$_{\rm p}$ [erg g$^{-1}$ K$^{-1}$] the specific heat capacity. 
$\partial$F = F$_{\uparrow, 2}$ - F$_{\downarrow, 2}$ - F$_{\uparrow, 1}$ +  F$_{\downarrow, 1}$  is the net radiative flux into the layer and $\partial$p = p$_{2}$ - p$_{1}$ is the pressure difference across the layer.

The AA method is only valid for weak scattering and does not consider the directional component of the scattering, assuming all scattering is in the exact forward direction (i.e. $g_{0}$ = 1).
$\delta$-M+ scaling of the albedo and optical depth for AA can be applied following \citet{Wiscombe1977} and \citet{Lin2018}.
We denote this scaled version as $\delta$-2AA/$\delta$-4AA \citep{Zhang2017}. 

\subsection{Extended Absorption Approximation}
\label{sec:EAA}

To somewhat take into account the directional component in the AA formulation, we explore a slight extension to the co-albedo, $\epsilon'$, following the hemispheric two-stream solution \citep[e.g.][]{Pierrehumbert2010,Heng2014, Heng2017}
\begin{equation}
    \epsilon' = \sqrt{(1 - \omega_{0})(1 - g_{0}\omega_{0})},
\end{equation}
where $g_{0}$ $\geq$ 0, is assumed for a forward or isotropic scattering component only.
The AA method is returned when $g_{0}$ = 1, with the regular definition of the co-albedo $\epsilon'$ = $\epsilon$.
When $g_{0}$ = 0 for isotropic scattering we more clearly see the effect of the altered co-albedo, this results in $\epsilon'$ = $\sqrt{\epsilon}$, increasing the co-albedo, in effect reducing the single scattering albedo.
This increase attempts to account for the non-preference in the scattering direction through reducing the transmission function.
Overall, this altered co-albedo attempts to relax the $g_{0}$ = 1 assumption in the AA approach in a simple manner, we name this alteration the `Extended Absorption Approximation' (EAA) or when $\delta$-M+ scaled $\delta$-nEAA for n streams. 
In Appendix \ref{app:AA_EAA} we explore specifically the differences between AA and EAA for our proposed tests.
Overall, we find EAA to be a general improvement over AA.

\section{Variational Iteration Method}
\label{sec:VIM}

\citet{Zhang2017} propose a variational iteration method (VIM) \citep[e.g.][]{He1999} solution to the longwave RT equation with approximate scattering.
In VIM, a differential equation is split into a linear and non-linear term.
An iterative scheme can then be used to improve the solution (the so called correction functional) by successive orders through applying a Lagrange multiplier to the previous solution.
Importantly, the VIM solution does not consider true multiple scattering across layers but only the scattering properties within a layer and their onward affect on adjacent layers.

Here we apply the four-stream variant of VIM outlined in \citet{Zhang2017}, however, we note the theory generalises to n-streams.
For brevity, we do not repeat the extensive derivation performed in \citet{Zhang2017}, only stating the main result of the derivation. 
The four-stream longwave radiative-transfer equation is approximated as \citep{Zhang2017}
\begin{multline}
    I_{\uparrow, 2}(\mu_{i}) = I_{\uparrow, 1}(\mu_{i})\mathcal{T}
    - d_{\rm i}(B_{2}\mathcal{T} - B_{1}) \\ 
    + \frac{\omega_{0}}{4\mu_{i}}\sum_{j=1}^{2}\{S_{\uparrow, j} - b_{i}[c_{j}\psi(\mu_{i},\mu_{j}) \\ +  c_{-j}\psi(\mu_{i},\mu_{-j})]  (B_{2}\mathcal{T} - B_{1})\},
\end{multline}
\begin{multline}
    I_{\downarrow, 2}(\mu_{-i}) = I_{\downarrow, 1}(\mu_{-i})\mathcal{T}
    - d_{\rm -i}(B_{2} - B_{1}\mathcal{T}) \\ 
    - \frac{\omega_{0}}{4\mu_{-i}}\sum_{j=1}^{2}\{S_{\downarrow, -j} - b_{-i}[c_{j}\psi(\mu_{-i},\mu_{j}) \\ +  c_{-j}\psi(\mu_{-i},\mu_{-j})]  (B_{2} - B_{1}\mathcal{T})\}.
\end{multline}
Here, i and j denote the positive cosine angles (i.e. $\mu_{i}$ = $\mu_{i}$)  and -i and -j the negative cosine angles (i.e. $\mu_{-i}$ = -$\mu_{i}$), with the coefficients b, c, d and $\psi$ given by
\begin{align} 
    b_{i} &= \frac{\mu_{i}}{\mu_{i}\beta + 1}, \\
    c_{i} &= \frac{\epsilon}{\mu_{i}\beta + \epsilon}, \\ 
    d_{i} &= \frac{\epsilon}{\mu_{i}\beta + 1}, \\
    \psi(\mu_{i},\mu_{j}) &= 1 + 3g_{0}\mu_{i}\mu_{j}.
\end{align}

The scattering coefficients S$_{\uparrow, j}$, S$_{\downarrow, j}$ are given by
\begin{multline}
S_{\uparrow, j} = \psi(\mu_{i},\mu_{j})\zeta_{i,-j}[I^{0}_{\downarrow, 1}(\mu_{-j}) - B_{1}c_{-j}](1 - e^{-\tau/\zeta_{i,-j}}) \\
+ \psi(\mu_{-i},\mu_{j})\zeta_{i,j}[I^{0}_{\uparrow, 2}(\mu_{j}) - B_{2}c_{j}] \times \\
[e^{-\epsilon\tau/\mu_{j}} - e^{-\tau/\mu_{i}}],
\end{multline}
and
\begin{multline}
S_{\downarrow, -j} = \psi(\mu_{-i},\mu_{-j})\zeta_{-i,j}[I^{0}_{\uparrow, 2}(\mu_{j}) - B_{2}c_{j}](1 - e^{-\tau/\zeta_{-i,j}}) \\
+ \psi(\mu_{i},\mu_{-j})\zeta_{-i,-j}[I^{0}_{\downarrow, 1}(\mu_{-j}) - B_{1}c_{-j}] \times \\
[e^{-\epsilon\tau/\mu_{j}} - e^{-\tau/\mu_{i}}],
\end{multline}
where
\begin{equation}
\zeta_{i,j} = \frac{\mu_{i}\mu_{j}}{\mu_{i}\epsilon - \mu_{j}}.
\end{equation}

In practice, VIM uses the (E)AA method to provide the zeroth order estimate of the intensity (here denoted by I$^{0}_{\uparrow,\downarrow}$) in the upward and downward directions. 
This zeroth order solution is then used to estimate the scattering coefficients S$_{\uparrow,\downarrow}$, with the final RT solution performed including the estimated scattering source function.
VIM therefore attempts to calculate a first order scattering `correction' to the zeroth order (E)AA method, following the iterative methodology of VIM.
In addition, in comparison to the \citet{Toon1989} two-stream method, the Henyey-Greenstein phase function is used for the scattering component rather than a hemispheric-mean approximation.
The main benefit of the VIM method is that when no scattering component is present in a layer, the scattering source function need not be calculated, substantially improving computational efficiency. 
Again, the $\delta$-M+ scaled versions of VIM are denoted $\delta$-nVIM.

\section{Testing the approximations}
\label{sec:tests}

In this section, we briefly describe the other RT methods used in this study to benchmark against the AA and VIM approximations.
We then perform tests on the computational speed, OLR and vertical heating rates for each method.

\subsection{Other RT methods}

The \citet{Toon1989} (hereafter Toon89) method applies a two-stream approximation to calculate the scattering source function assuming a hemispheric mean scattering phase function.
Then, the scattering source function is added to thermal component in the source function technique to find the intensity.
Any number of streams can be used for the source function technique part of the calculation, but the scattering component is always assumed as the hemispheric mean two-stream solution.
Here, we use four-stream version for the source function technique part of the calculation for a fairer comparison to the other four-stream approximate methods used in this study.

We note \citet{Marley2021}, which also used Toon89, have applied 12-streams for the source functions part of the method in their brown dwarf RCE modelling.
Toon89 has also been applied to numerous exoplanet GCM models \citep[e.g.][]{Showman2009,Roman2021,Wolf2022,Lee2023} and retrieval/RCE models \citep[e.g.][]{Line2013, Burningham2017,Batalha2019} with various number of source function streams.
However, the two-stream derived hemispheric mean scattering source function remains the same for all these studies.

A specific version of the discrete ordinance method code DISORT \citep{Stamnes1988,Stamnes2000} optimised for two-streams (DISORT-twostr) is publicly available\footnote{\url{http://www.rtatmocn.com/disort/}} \citep{Kylling1995}.
Some brown dwarf and exoplanet GCM studies have used this code as the RT driver \citep[e.g.][]{Tan2021}, mostly with application coupling to the MITgcm \citep{Komacek2022}.
We use the DISORT code with 64 streams as our reference solutions to compare methods.

\subsection{Grey speed tests}

\begin{table}[]
    \centering
    \begin{tabular}{c|c|c|c|c|c} \hline \hline
      Method & $\delta$-2EAA & $\delta$-4EAA  & $\delta$-4VIM & Toon89 & DISORT-twostr  \\ \hline
      $\omega_{0}$ = 0 & 1.0 & 1.15 & 1.15 & 3.34 & 4.91  \\
      $\omega_{0}$ = 0.5 & 1.0 & 1.18 & 3.68 & 3.49 & 5.43  \\ \hline \hline
    \end{tabular}
    \caption{Relative speed scaled to $\delta$-2EAA of each of the longwave algorithms for the grey atmosphere speed test.}
    \label{tab:speed}
\end{table}

For a fair test of the pure RT component of each method, we time the longwave algorithm for each method for a grey opacity atmospheric model.
This is ideal to test the direct speed properties as it greatly simplifies the RT problem and only one pass of the longwave subroutine is required per iteration.

We use the two-stream AA as the baseline, adding the four-stream VIM, four-stream Toon89 method and the two-stream DISORT version to the speed test.
We use a \citet{Guillot2010} analytical radiative-equilibrium grey opacity profile, with first test without and scattering component and the second test with a constant single scattering albedo of 0.5 and asymmetry factor of 0.5.
Table \ref{tab:speed} shows the results of this speed test.
The main result of this test is that the speed of the $\delta$-4VIM method is on par with the Toon89 method when every layer contains a scattering component, but speed is much improved in $\delta$-4VIM method the when no scattering components are present.

\subsection{Outgoing longwave radiation fluxes}

\begin{figure}
    \centering
    \includegraphics[width=0.49\textwidth]{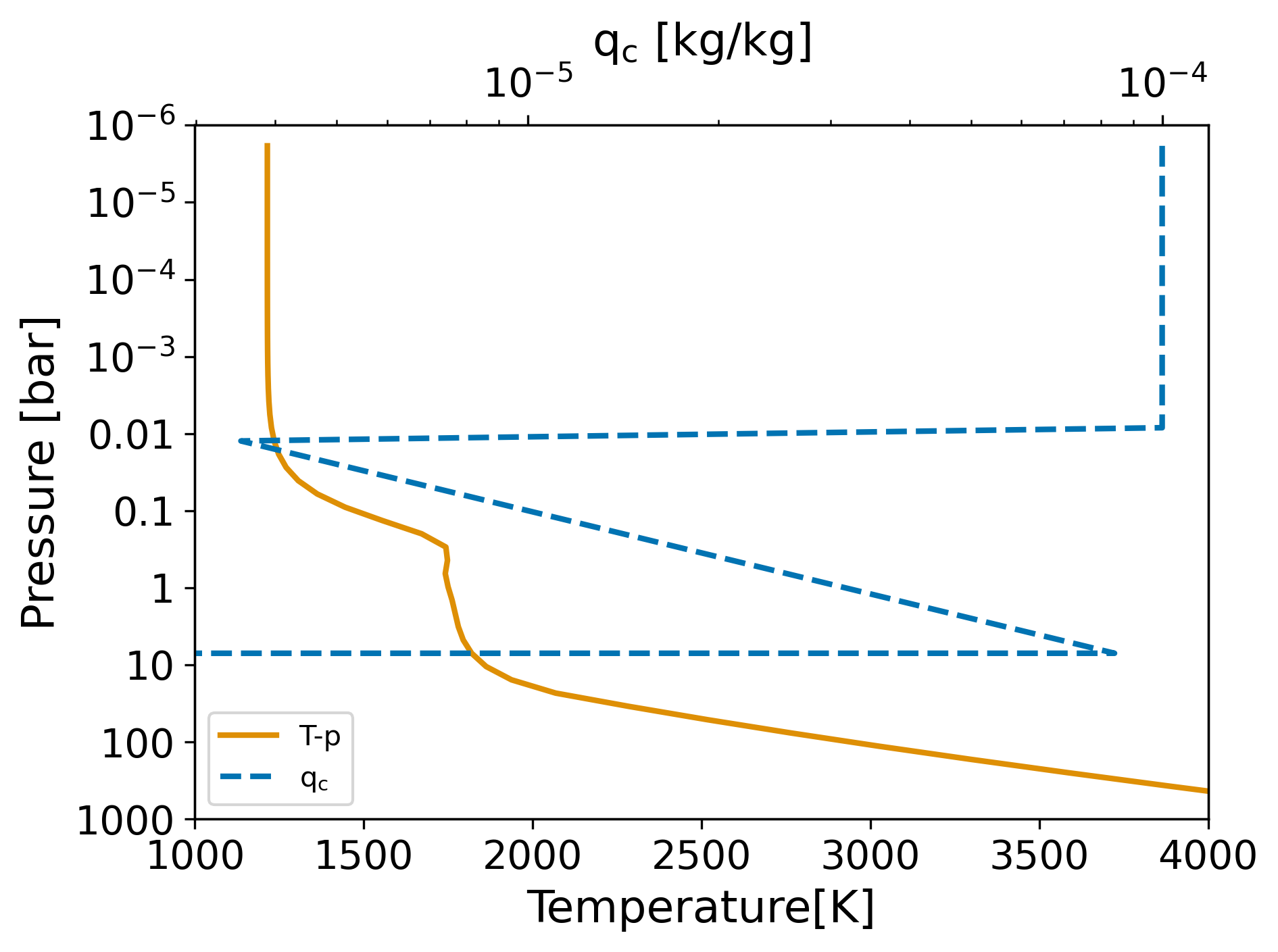}
    \caption{HD 189733b global averaged T-p profile (orange solid line) produced from the picket-fence scheme and condensed mass fraction q$_{\rm c}$ (blue dashed line) profiles for the parameterised deep and extended cloud layers used in the RT tests.}
    \label{fig:T-p}
\end{figure}

\begin{figure*}
    \centering
    \includegraphics[width=0.49\textwidth]{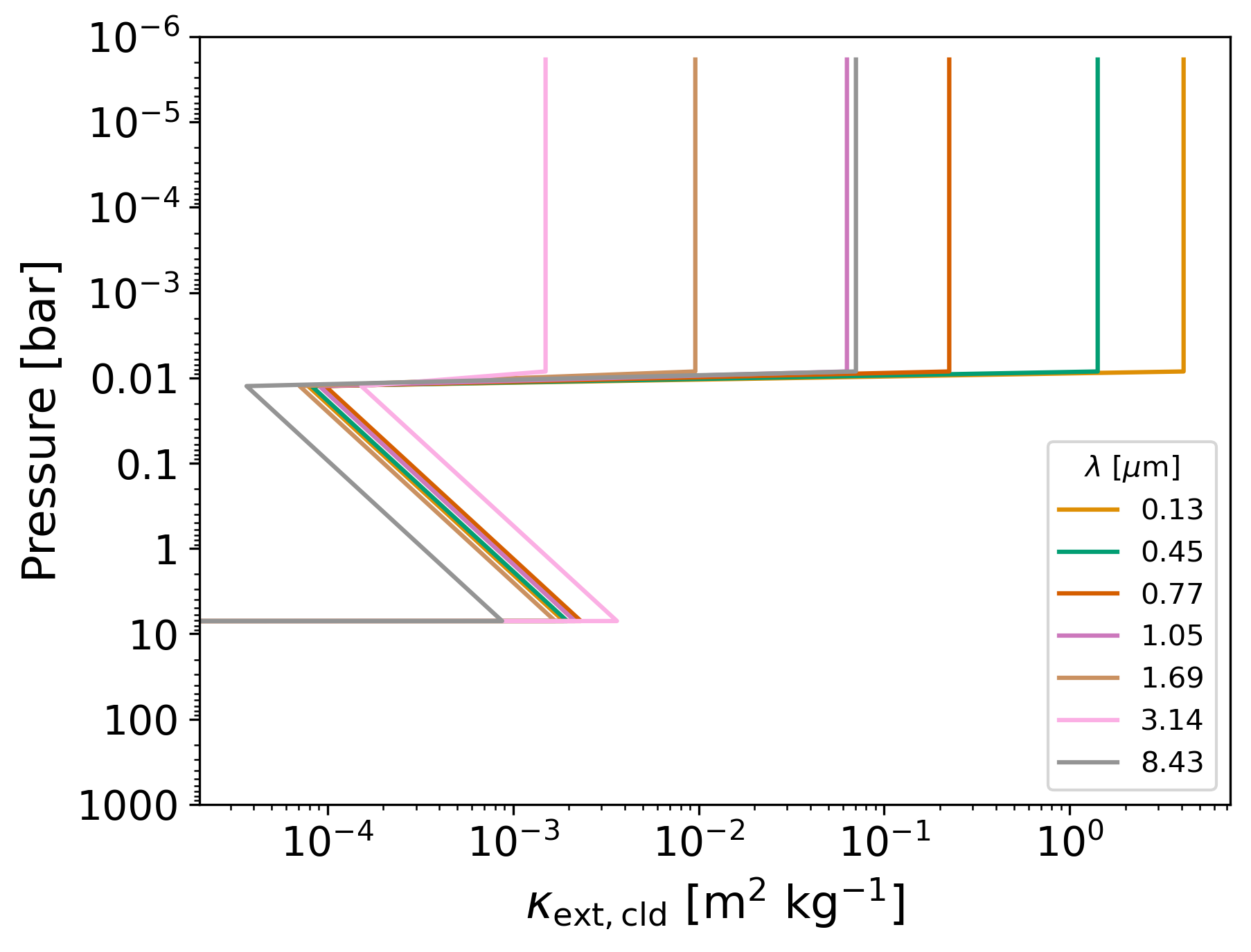}
    \includegraphics[width=0.49\textwidth]{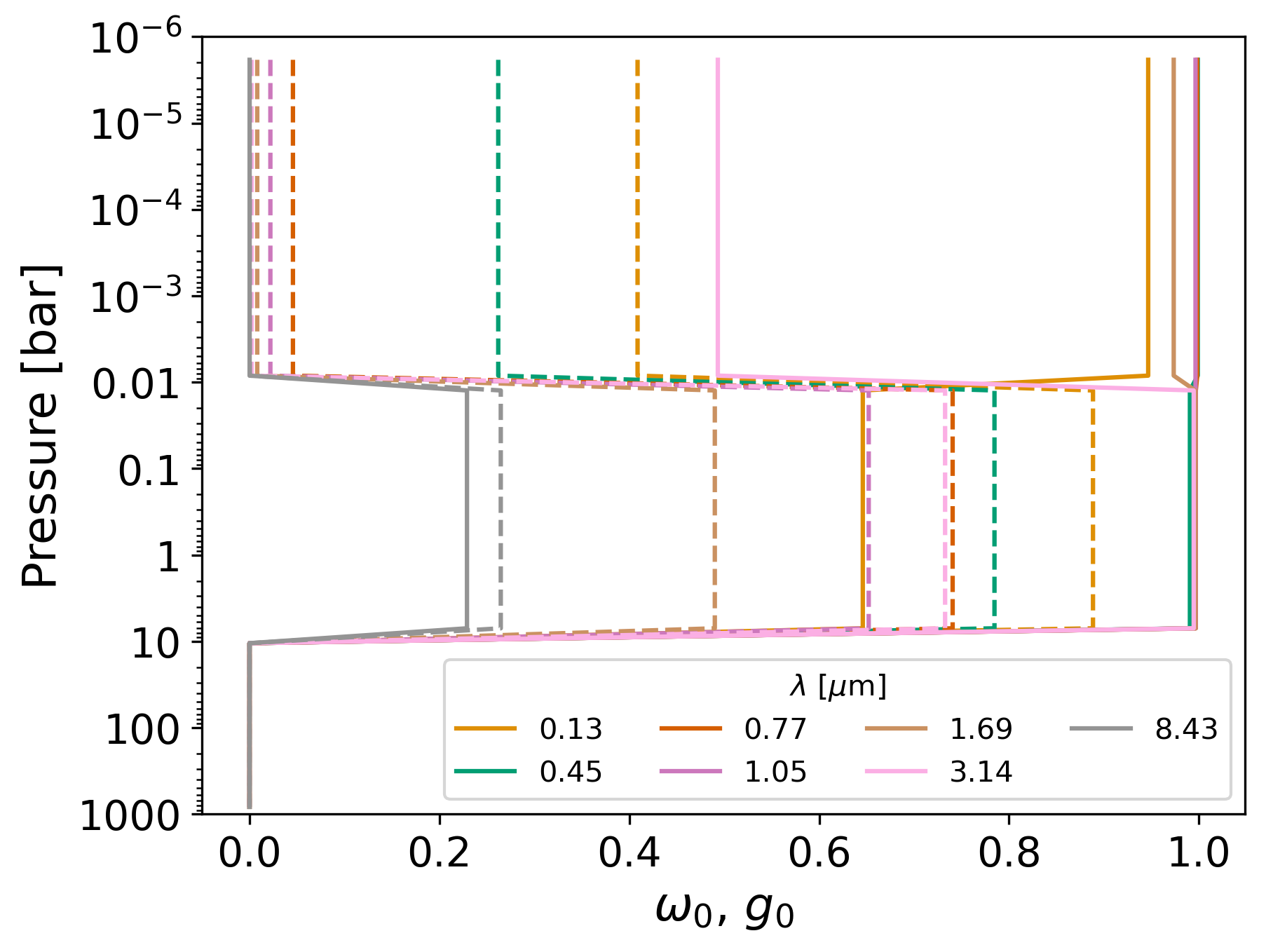}
    \caption{Left: cloud extinction opacity, $\kappa_{\rm ext, cld}$, with pressure for the combined upper and lower cloud layer case for several wavelength bands. 
    Right: single scattering albedo, $\omega_{0}$ (solid lines), and asymmetry parameter, $g_{0}$ (dashed lines), with pressure for several wavelength bands for the combined cloud case.}
    \label{fig:prop}
\end{figure*}

\begin{table}
    \centering
    \begin{tabular}{c|c|c|c|c} \hline \hline
    Scenario & $\delta$-4EAA & $\delta$-4VIM & Toon89 & DISORT-twostr \\ \hline
    Cloud free & 0.73 & 0.60 & 0.33 &  3.67 \\
    Deep cloud & 0.37 & 0.98 & 0.36 & 3.99  \\
    Extended cloud & -2.05 & 0.89 & 0.30 & 4.35 \\
    Both clouds & -2.33 & 1.21 & 0.36 & 4.66  \\ \hline
    \end{tabular}
    \caption{Relative percentage error on the total OLR for each method and scenario compared to the reference 64 stream DISORT results.}
    \label{tab:olr}
\end{table}

In this section, we test the accuracy of each scheme in producing OLR fluxes under cloud free and cloudy scenarios.
We follow a similar methodology to \citet{Amundsen2014}, where a representative HD 209458b hot Jupiter T-p profile from the \citet{Heng2011} polynomial fit to the \citet{Iro2005} simulations was used to test the radiative-transfer schemes.
Here, we use a global averaged HD 189733b profile derived from the \citet{Parmentier2015} analytical picket-fence scheme as our test temperature-pressure (T-p) profile.
Figure \ref{fig:T-p} shows this T-p profile.
A deep boundary pressure of 1000 bar is chosen provide a large optical depth region to avoid undue boundary condition influences on the upper atmosphere.
We use 54 layers, emulating a typical GCM vertical resolution. 
We use the same hybrid-sigma grid pressure spacing as used in the hot Jupiter Exo-FMS simulations \citep[e.g.][]{Lee2021}.
Strong evidence of cloud coverage on HD 189733b is seen in contemporary observational data \citep[e.g.][]{Sing2016}, making the HD 189733b hot Jupiter parameter regime ideal for testing the longwave schemes.

For the gas phase opacity, we use a 34 band scheme the same as \citet{Amundsen2014} with pre-mixed correlated-k tables \citep{Amundsen2017}, assuming mixing ratios at chemical equilibrium. 
These tables use a 4+4 k-coefficient scheme the same as \citet{Kataria2013} and \citet{Marley2021} and have been previously used in hot Jupiter simulations using Exo-FMS \citep[e.g.][]{Lee2021}.
Pre-mixed tables such as these are in common use in GCMs of exoplanet atmospheres \citep[e.g.][]{Showman2009,Schneider2022}

For the cloud opacities, we apply Mie theory using a Fortran conversion of LX-MIE code from \citet{Kitzmann2018}.
We convert a given cloud mass mixing ratio, q$_{\rm c}$ [kg/kg], to a total cloud particle number density, assuming a mean cloud particle size and standard deviation, following a log-normal distribution \citep[e.g.][]{Komacek2022}.
This results in realistic cloud opacities, single scattering albedo and asymmetry parameters across all the bands.

We investigate three cloud scenarios, a deep cloud case, an upper extended cloud case and a combination deep and extended case.
For the deep cloud case, we parameterise a cloud base at 10 bar and cloud top at 0.01 bar, using a mean particle size of 2$\mu$m and standard deviation of $\sigma$ = 1.
The q$_{\rm c}$ is given a maximum value at 10 bar of 10$^{-4}$ kg/kg, with a power law drop off with gradient = 0.5 to the cloud top pressure level.
This mimics a compact deep cloud layer scenario, such as those seen in \citet{Ackerman2001} models of sub-stellar objects \citep[e.g.][]{Gao2018}, when mixing is slow and sedimentation efficient.

For the extended cloud case, the cloud base is at 0.01 bar and cloud top at 10$^{-6}$ bar, using a mean particle size of 0.1$\mu$m and standard deviation of $\sigma$ = 0.1.
We use a constant q$_{\rm c}$ of 10$^{-4}$ kg/kg.
This intends to represent a well mixed and lofted upper atmosphere extended cloud, commonly seen for models that contain strong mixing or those seen in cloud studies using GCM model output \citep[e.g.][]{Gao2018, Powell2018, Lee2023}.

For both cloud layers, we assume a MgSiO$_{3}$ cloud composition, taking the optical constants of amorphous MgSiO$_{3}$ from the \citet{Kitzmann2018} collection.
Figure \ref{fig:T-p} shows the q$_{\rm c}$ structure of the two cloud parameterisations.
After the calculation of the cloud opacity, several wavelength bands exhibit high single scattering albedos coupled with high asymmetry factors (e.g. $\omega_{0}$ $>$ 0.8, $g_{0}$ $>$ 0.8) inside both cloud regions, which is an ideal case for testing the scattering properties of each method.
For illustration, Figure \ref{fig:prop} shows several wavelength bands and their cloud opacity, single scattering albedo and asymmetry parameter with pressure.

In Table \ref{tab:olr}, we present the total OLR error for each method compared to the reference 64-stream DISORT calculations.
Here we see the EAA method varies strongly depending on the cloud component, with the deep clouds more accurately represented than the extended cloud, or even cloud free profile.
VIM performs similarly across all tests, with an error of around $\sim$1\%.
Toon89 fares the best, with OLR errors of only $\sim$0.3\% compared to the reference DISORT calculations, consistent across all tests.
The DISORT-twostr methods produces the largest errors at around $\sim$4\%.

\subsection{Heating rates}

\begin{figure*}
    \centering
    \includegraphics[width=0.41\textwidth]{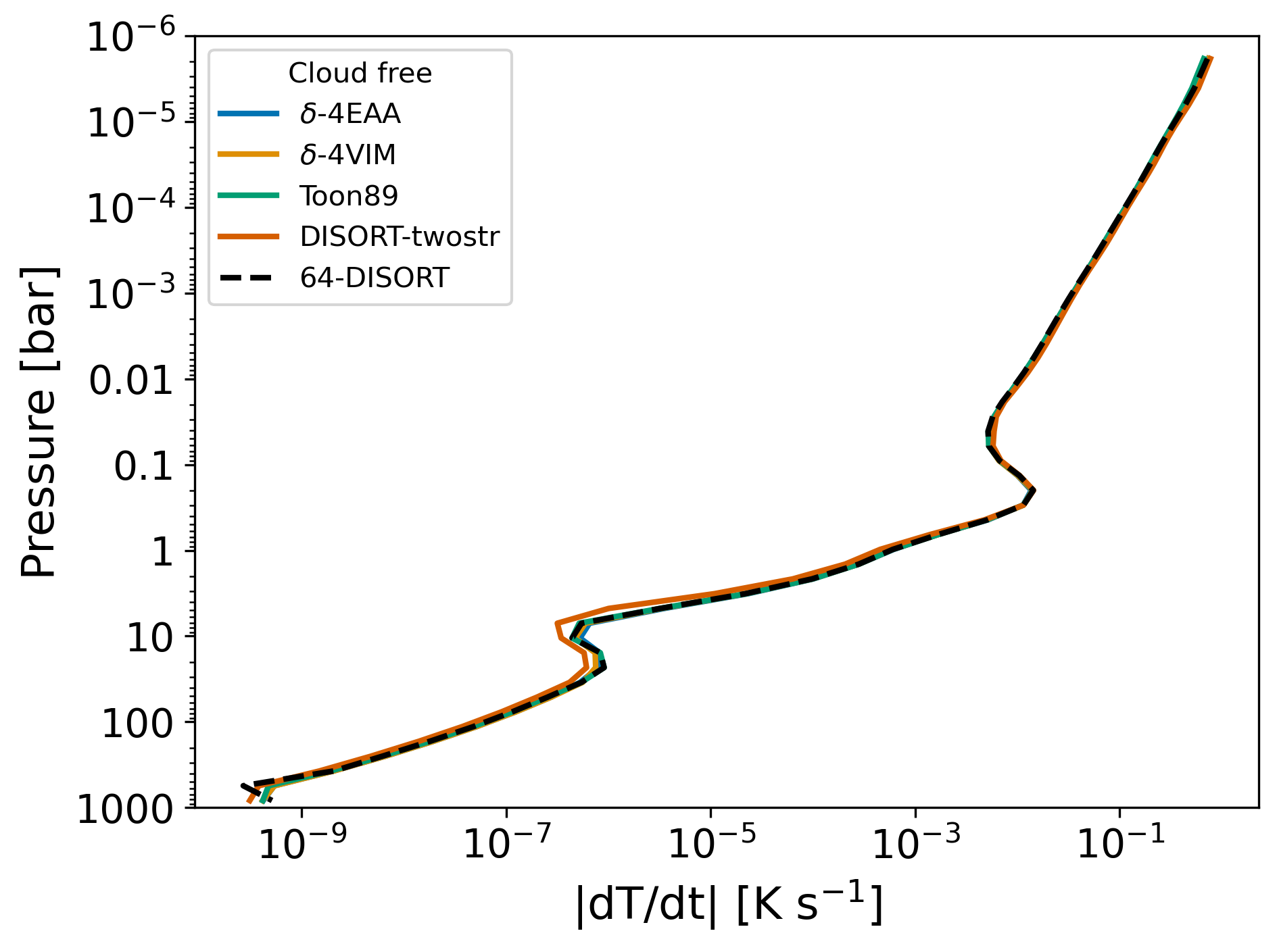}
    \includegraphics[width=0.41\textwidth]{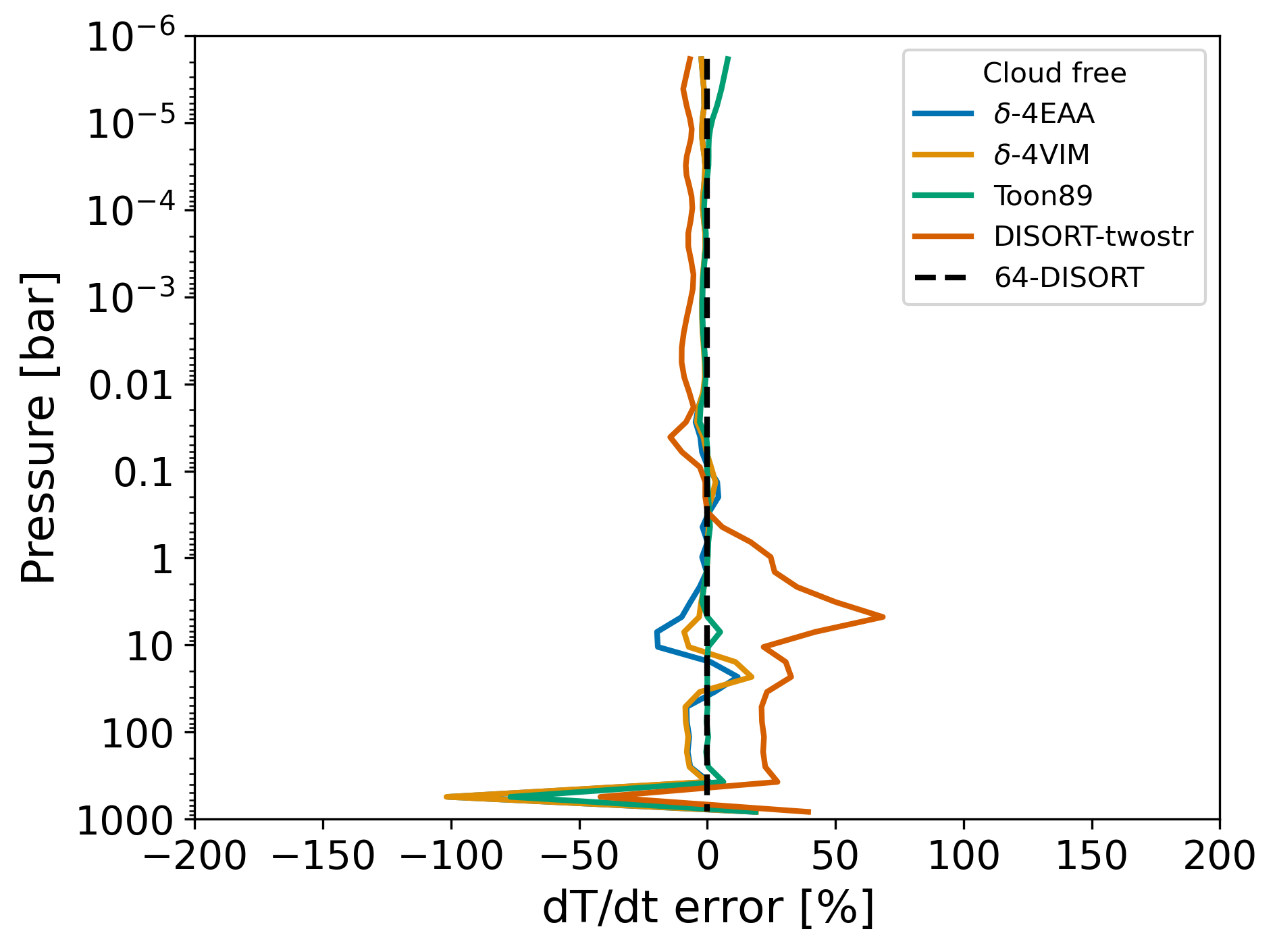}
    \includegraphics[width=0.41\textwidth]{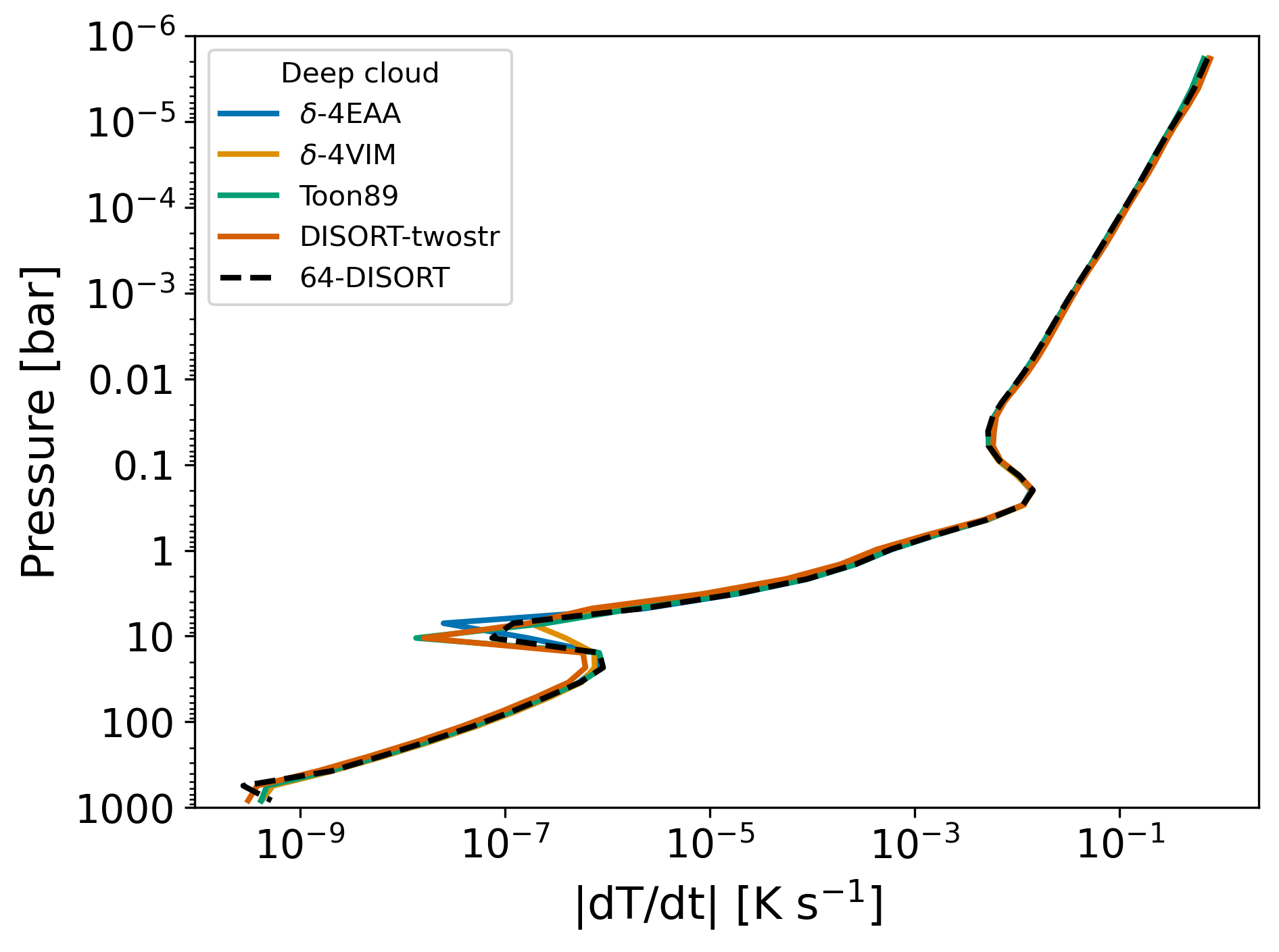}
    \includegraphics[width=0.41\textwidth]{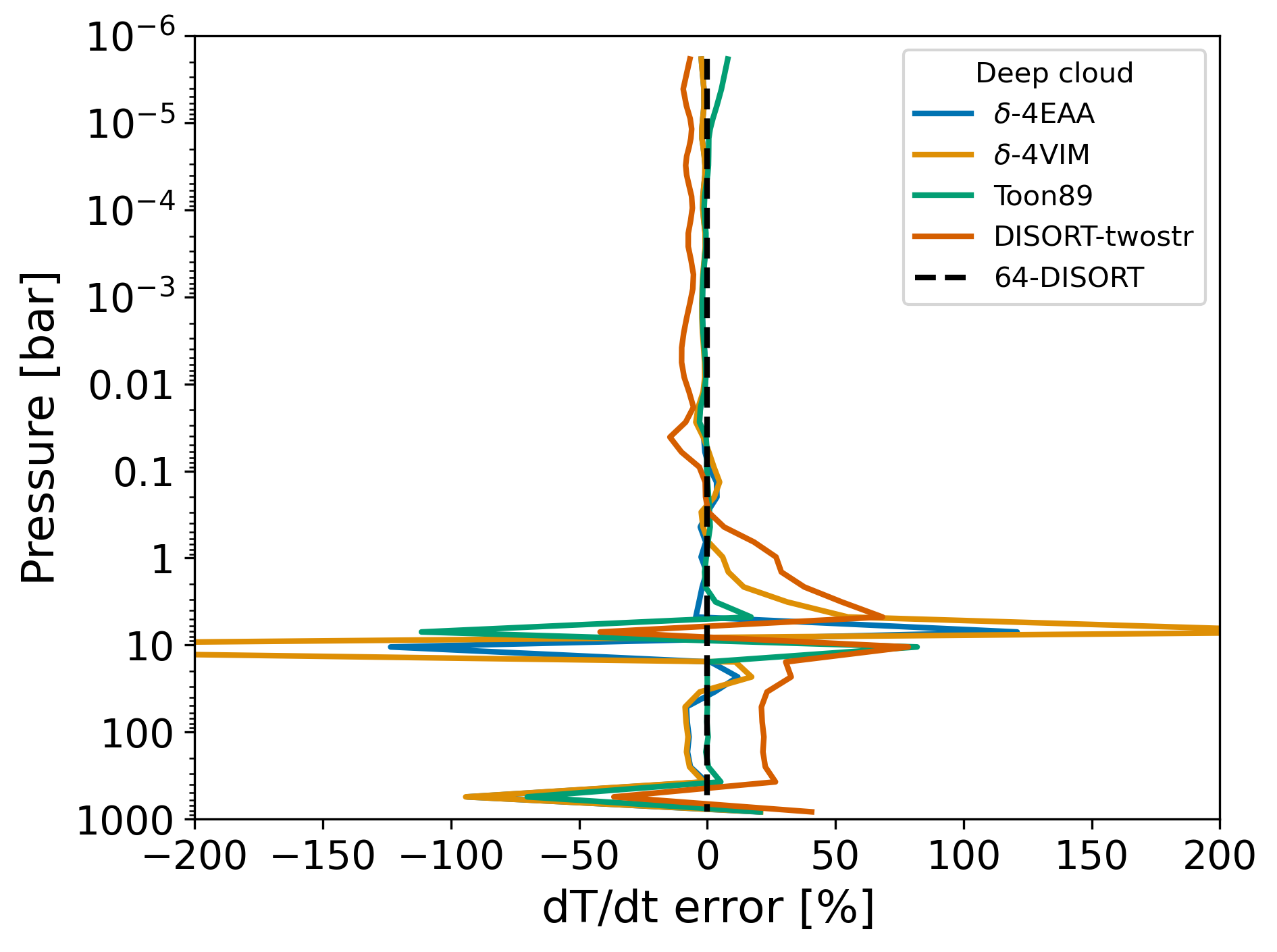}
    \includegraphics[width=0.41\textwidth]{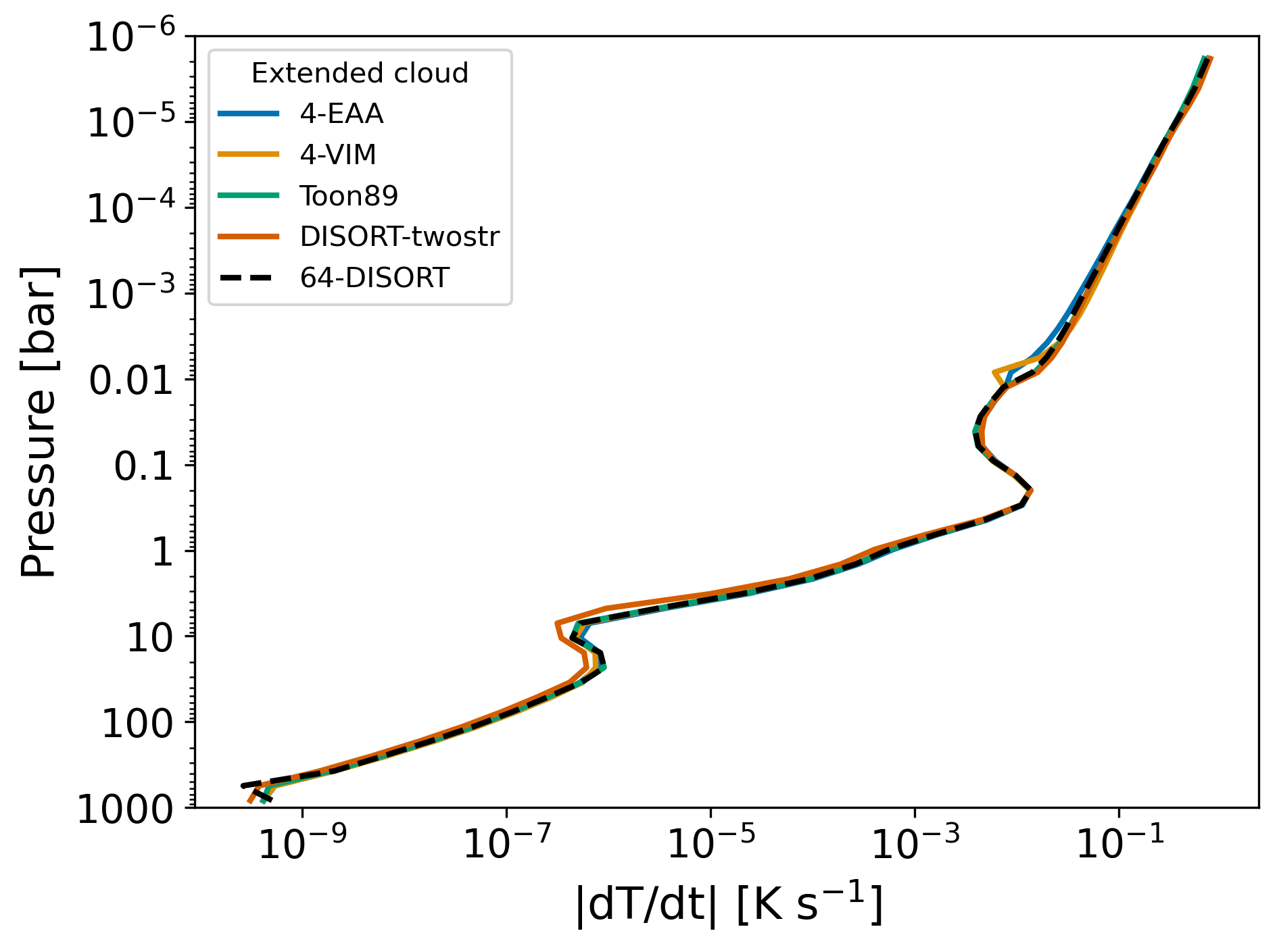}
     \includegraphics[width=0.41\textwidth]{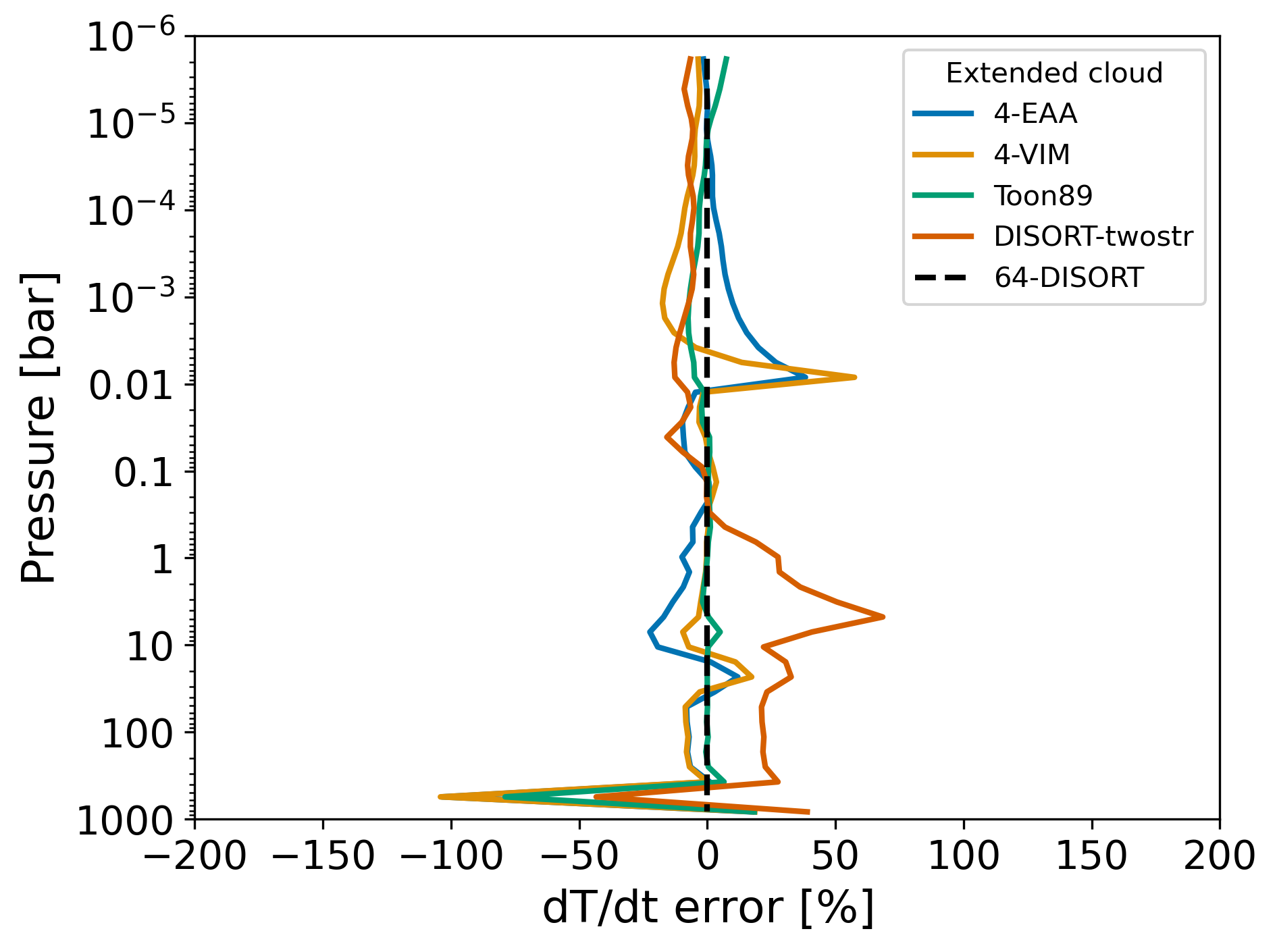}
    \includegraphics[width=0.41\textwidth]{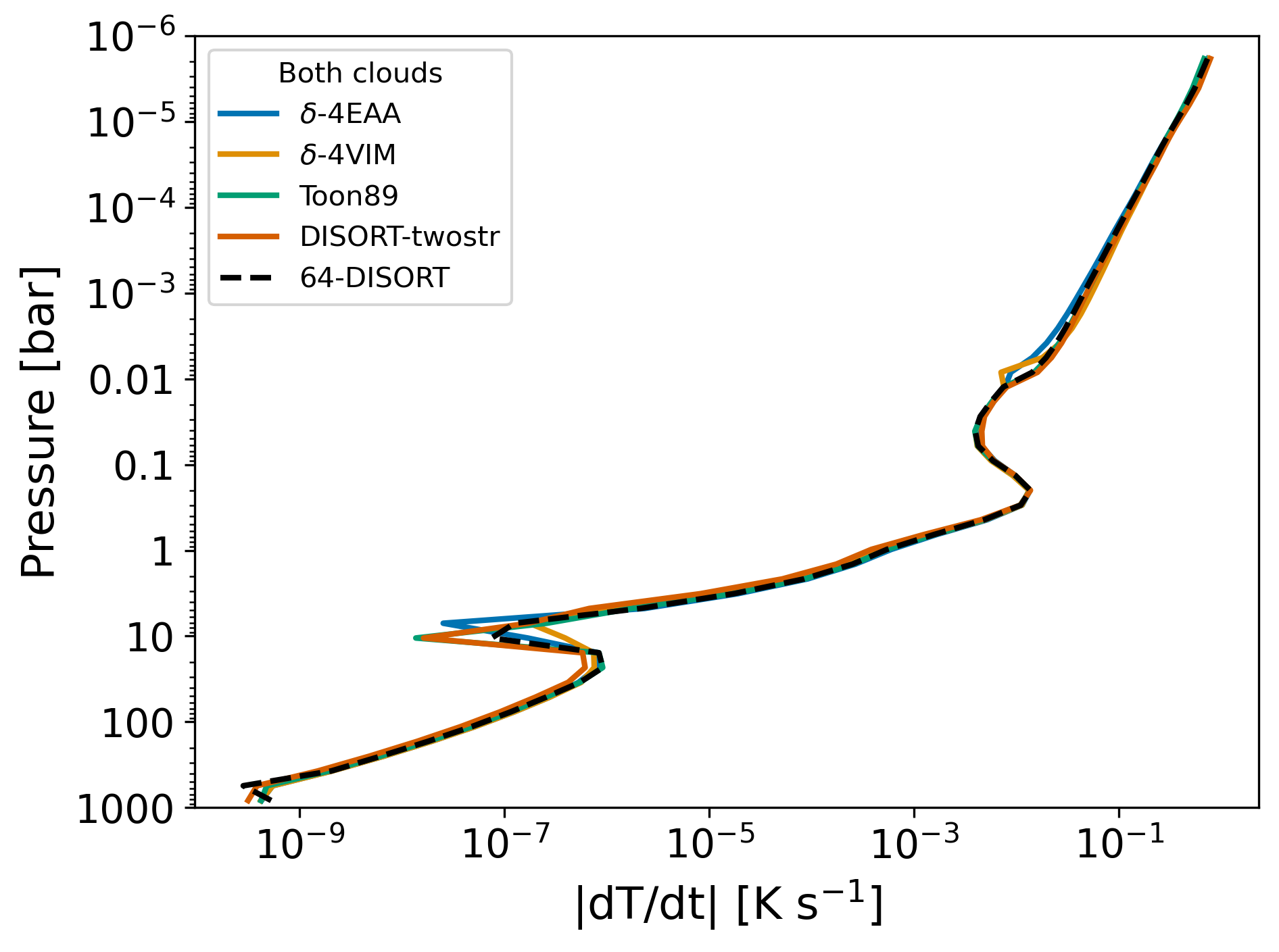}
    \includegraphics[width=0.41\textwidth]{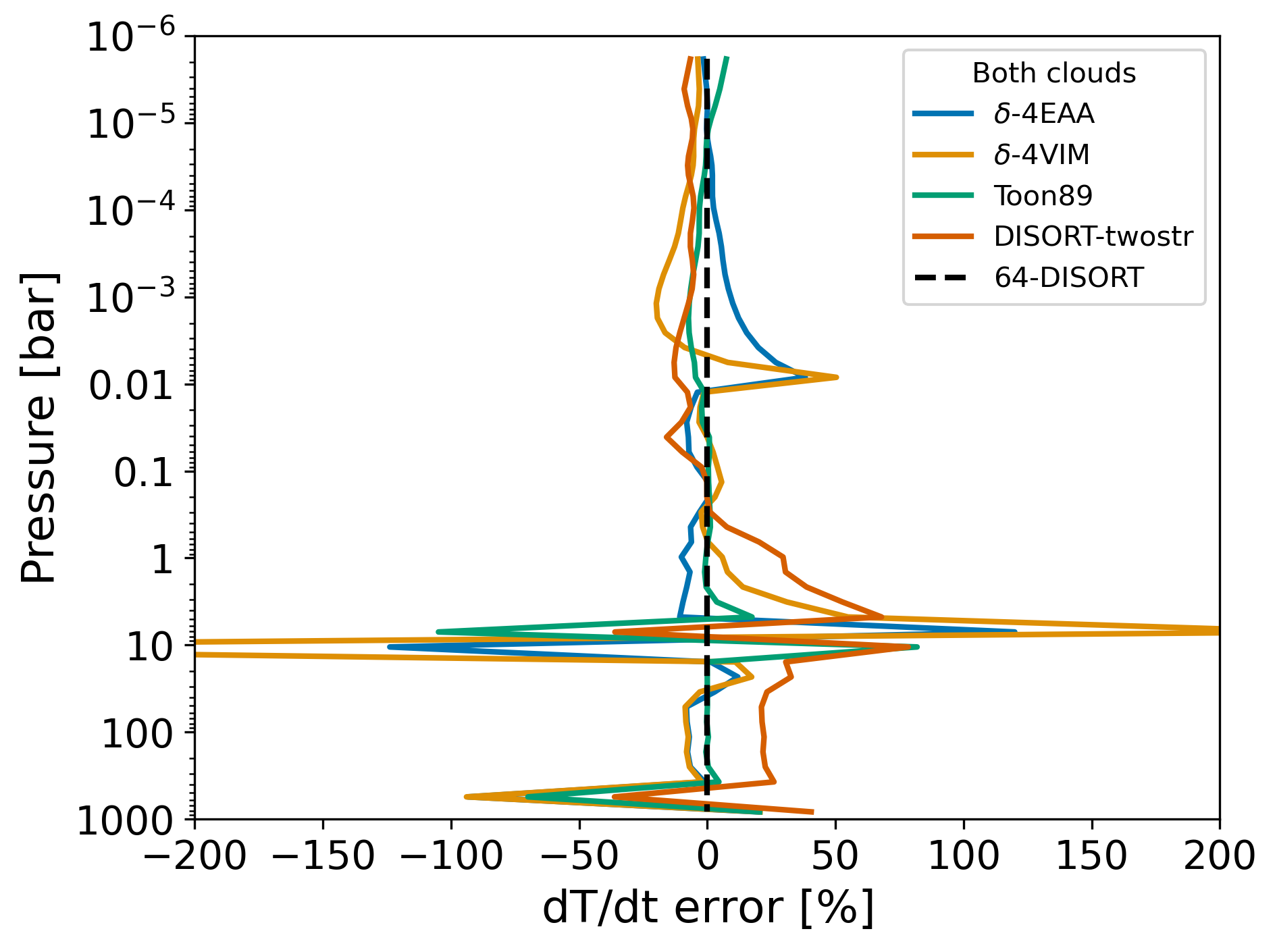}
    \caption{Absolute value of the vertical heating rates (left) and percent error in the heating rates (right) for each of the test cases. 
    Each method is shown as a coloured solid line.}
    \label{fig:hrate}
\end{figure*}

In this section, we test the accuracy of each scheme in producing vertical heating fluxes under cloud free and cloudy scenarios.
We use the same T-p, opacity and cloud component set-up as the previous section.
Figure \ref{fig:hrate} presents the results of the tests, showing both the absolute value of the heating rates produced by each method and the error relative to the 64-stream DISORT reference solution.
For the cloud free case, all methods produce very reliable results, with DISORT-twostr producing the largest errors of about 50\% in deeper regions.

For the deep cloud case we start to see where the main differences between the methods arise, where all methods produce significant errors near the base of cloud layer where the opacity of the cloud is strongest.
Interestingly, the Toon89 and DISORT-twostr methods produce heating locations similar to DISORT while the EAA and VIM methods results in a slightly lower pressure heating spike due to the cloud layer.
The large errors in the AA and VIM method at these pressures reflect this difference in heating level.
This suggests that multiple-scattering may be playing an important role here, allowing a heating component below the cloud base not captured by AA or VIM.
However, the multiple-scattering methods still show significant error at the cloud base, with up to 50\% errors compared to the DISORT reference solution.
Above the deep cloud base, errors are highly comparable between each method, suggesting this effect does not persist through the cloud layer.

For the extended cloud case, the Toon89 and DISORT-twostr compare much better than EAA and VIM to the reference solution, with EAA and VIM producing a greater heating component at the cloud base than the Toon89 and DISORT-twostr methods.
When comparing the results of both cloud layers, both errors in each scenario are generally added together, showing similar characteristics for each test separately. 

\section{Discussion}
\label{sec:disc}

Our OLR tests showed that the most consistent methods were Toon89 at $\sim$0.3\% error and VIM at $\sim$1\% error across all tests.
However, the computational time to accuracy trade-off must be taken into account for GCM and retrieval modelling efforts.
From the OLR tests, VIM is promising as in the best case scenario VIM is approximately 3x faster than Toon89 while at worst (a profile with all layers containing a scattering component) is on-par with the Toon89 computational time.
We therefore suggest that EAA and VIM show strong promise for improving the speed of emission spectra retrieval modelling where retrieving vertical cloud properties is required.
This is possibly an improvement for 2D retrieval efforts where hemispheres of cloud free and cloudy profiles are mixed.
EAA and VIM may be most useful for parameterised cloud profiles or grey clouds that have a specified single scattering albedo that is not too large (we suggest $\omega_{0}$ $\lesssim$ 0.8).
Usefully, both EAA and VIM can be used to retrieve an asymmetry parameter for a parameterised cloud layer.

Our vertical heating rate tests shows the strongest promise for approximate methods, as the errors are very comparable to the Toon89 and DISORT-twostr methods.
This suggests EAA and VIM to be very suitable for GCM modelling where RT efficiency is paramount.
The computational efficiency of VIM will also benefit from the vertical and hemispheric inhomogeneous nature of clouds on hot Jupiter. 

In \citet{Rooney2024}, the Toon89 method was compared to a spherical harmonic (SH) method with source function technique.
They found that the Toon89 method produced spectral OLR differences of $\sim$20\% or more at certain wavelengths compared to their DISORT reference calculations for their cloudy brown dwarf examples.
However, here we report a much closer agreement in OLR $<$ 1\% between the Toon89 and DISORT methods.
We suggest a few reasons for this discrepancy
\begin{itemize}
    \item The larger gravity used in the \citet{Rooney2024} brown dwarf models, making the OLR more sensitive to strong scattering components deeper in the atmosphere.
    \item The consistently large single scattering albedo and asymmetry factors produced by their cloud model across a wide infrared wavelength regime.
\end{itemize}
We suspect these two factors, as well as the assumption of a hemispheric mean phase function in the Toon89 method compared to the HG function in the SH method, are responsible for the much larger errors seen in \citet{Rooney2024}.
Owing to this, the EAA and VIM methods will need to be separately tested for brown dwarf conditions similar to \citet{Rooney2024}, though we speculate that since a lot of the errors in the Toon89 method stem from the use of the hemispheric-mean phase approximation \citep{Rooney2024} that VIM may be more accurate than Toon89 in some cases, but probably not as accurate as the SH method itself.

The VIM method is also known to be less accurate for smaller optical depth layers with large single scattering albedos \citep{Zhang2017}, {such as high altitude Earth clouds, this is primarily due to the zeroth order solution using AA which also performs worse at these conditions \citep{Zhang2017}. 
This can be seen in Figure \ref{fig:hrate}, where the EAA and VIM method produce a larger error for the low pressure extended cloud case compared the Toon89 and DISORT-twostr methods.
We therefore suggest this be taken into account when considering EAA or VIM for heating rate purposes, for example, this error may be acceptable for GCM modelling but not adequate for RCE modelling, where an inaccurate equilibrium profile would more greatly affect the interpretation of results.

Due to its application in Earth atmospheric modelling, the VIM method may be applied to Mars CO$_{2}$ clouds and their affect on the Mars climate \citep[e.g.][]{Forget1997, Kitzmann2016, Heng2017}.
However, as an approximate method, VIM is likely to not be as accurate as the DISORT simulations performed in \citet{Kitzmann2016}.
As stated in \citet{Zhang2017}, low optical depths with large albedo is a parameter regime where VIM can be in significant error, which is the case for the Mars CO$_{2}$ cloud problem.

\section{Summary and Conclusions} 
\label{sec:conc}

In this brief study, we investigated two approximations for calculating internal atmospheric  (longwave) radiation with a scattering component in a hot Jupiter context, the (extended) absorption approximation (EAA) and the variational iteration method (VIM).

For retrieval modelling, EAA and VIM may be most useful for parameterised cloud profiles or grey clouds that have a specified single scattering albedo that is not too large.
In the JWST era, where the number of wavelength points required for calculation is steadily increasing, and as retrieving specific cloud properties becomes more important, convergence runtimes of retrieval models are also on the rise.
EAA and VIM offer a simple way to speed up the forward modelling process without too much sacrifice in accuracy.

For gas giant exoplanet GCMs, VIM offers a quick and simple method to take into account cloud particle scattering in the longwave, with comparable heating rate errors to other well used models in the field.
We therefore suggest VIM is a highly promising method for use in exoplanet GCMs.

Due to the generalised nature of the RT methods used here, our findings are likely applicable to atmospheres across the exoplanet spectrum, from ultra hot Jupiters to warm Neptunes and rocky planets.
However, we caution high gravity, strong scattering cloudy brown dwarf conditions such as those in \citet{Rooney2024} will need to be tested further before AAE and VIM can be recommended for brown dwarf modelling efforts.

\begin{acknowledgments}
E.K.H. Lee is supported by the SNSF Ambizione Fellowship grant (\#193448).
\end{acknowledgments}

\vspace{5mm}

\software{DISORT and DISORT-twostr: \url{http://www.rtatmocn.com/disort/}. 
Exo-FMS column emulation codes: \url{https://github.com/ELeeAstro}. 
Fortran 90 version of LX-MIE: \url{https://github.com/ELeeAstro/gCMCRT/blob/main/src_optools_V2/lxmie_mod.f90}}

\bibliography{bib}{}

\begin{thebibliography}{}
\expandafter\ifx\csname natexlab\endcsname\relax\def\natexlab#1{#1}\fi
\providecommand{\url}[1]{\href{#1}{#1}}
\providecommand{\dodoi}[1]{doi:~\href{http://doi.org/#1}{\nolinkurl{#1}}}
\providecommand{\doeprint}[1]{\href{http://ascl.net/#1}{\nolinkurl{http://ascl.net/#1}}}
\providecommand{\doarXiv}[1]{\href{https://arxiv.org/abs/#1}{\nolinkurl{https://arxiv.org/abs/#1}}}

\bibitem[{{Ackerman} \& {Marley}(2001)}]{Ackerman2001}
{Ackerman}, A.~S., \& {Marley}, M.~S. 2001, \apj, 556, 872,
  \dodoi{10.1086/321540}

\bibitem[{{Amundsen} {et~al.}(2014){Amundsen}, {Baraffe}, {Tremblin},
  {Manners}, {Hayek}, {Mayne}, \& {Acreman}}]{Amundsen2014}
{Amundsen}, D.~S., {Baraffe}, I., {Tremblin}, P., {et~al.} 2014, \aap, 564,
  A59, \dodoi{10.1051/0004-6361/201323169}

\bibitem[{{Amundsen} {et~al.}(2017){Amundsen}, {Tremblin}, {Manners},
  {Baraffe}, \& {Mayne}}]{Amundsen2017}
{Amundsen}, D.~S., {Tremblin}, P., {Manners}, J., {Baraffe}, I., \& {Mayne},
  N.~J. 2017, \aap, 598, A97, \dodoi{10.1051/0004-6361/201629322}

\bibitem[{{Arcangeli} {et~al.}(2019){Arcangeli}, {D{\'e}sert}, {Parmentier},
  {Stevenson}, {Bean}, {Line}, {Kreidberg}, {Fortney}, \&
  {Showman}}]{Arcangeli2019}
{Arcangeli}, J., {D{\'e}sert}, J.-M., {Parmentier}, V., {et~al.} 2019, \aap,
  625, A136, \dodoi{10.1051/0004-6361/201834891}

\bibitem[{{Batalha} {et~al.}(2019){Batalha}, {Marley}, {Lewis}, \&
  {Fortney}}]{Batalha2019}
{Batalha}, N.~E., {Marley}, M.~S., {Lewis}, N.~K., \& {Fortney}, J.~J. 2019,
  \apj, 878, 70, \dodoi{10.3847/1538-4357/ab1b51}

\bibitem[{{Burningham} {et~al.}(2017){Burningham}, {Marley}, {Line}, {Lupu},
  {Visscher}, {Morley}, {Saumon}, \& {Freedman}}]{Burningham2017}
{Burningham}, B., {Marley}, M.~S., {Line}, M.~R., {et~al.} 2017, \mnras, 470,
  1177, \dodoi{10.1093/mnras/stx1246}

\bibitem[{{Changeat} {et~al.}(2021){Changeat}, {Al-Refaie}, {Edwards},
  {Waldmann}, \& {Tinetti}}]{Changeat2021}
{Changeat}, Q., {Al-Refaie}, A.~F., {Edwards}, B., {Waldmann}, I.~P., \&
  {Tinetti}, G. 2021, \apj, 913, 73, \dodoi{10.3847/1538-4357/abf2bb}

\bibitem[{{Coulombe} {et~al.}(2023){Coulombe}, {Benneke}, {Challener},
  {Piette}, {Wiser}, {Mansfield}, {MacDonald}, {Beltz}, {Feinstein}, {Radica},
  {Savel}, {Dos Santos}, {Bean}, {Parmentier}, {Wong}, {Rauscher}, {Komacek},
  {Kempton}, {Tan}, {Hammond}, {Lewis}, {Line}, {Lee}, {Shivkumar},
  {Crossfield}, {Nixon}, {Rackham}, {Wakeford}, {Welbanks}, {Zhang}, {Batalha},
  {Berta-Thompson}, {Changeat}, {D{\'e}sert}, {Espinoza}, {Goyal},
  {Harrington}, {Knutson}, {Kreidberg}, {L{\'o}pez-Morales}, {Shporer}, {Sing},
  {Stevenson}, {Aggarwal}, {Ahrer}, {Alam}, {Bell}, {Blecic}, {Caceres},
  {Carter}, {Casewell}, {Crouzet}, {Cubillos}, {Decin}, {Fortney}, {Gibson},
  {Heng}, {Henning}, {Iro}, {Kendrew}, {Lagage}, {Leconte}, {Lendl},
  {Lothringer}, {Mancini}, {Mikal-Evans}, {Molaverdikhani}, {Nikolov}, {Ohno},
  {Palle}, {Piaulet}, {Redfield}, {Roy}, {Tsai}, {Venot}, \&
  {Wheatley}}]{Coulombe2023}
{Coulombe}, L.-P., {Benneke}, B., {Challener}, R., {et~al.} 2023, \nat, 620,
  292, \dodoi{10.1038/s41586-023-06230-1}

\bibitem[{{Crouzet} {et~al.}(2014){Crouzet}, {McCullough}, {Deming}, \&
  {Madhusudhan}}]{Crouzet2014}
{Crouzet}, N., {McCullough}, P.~R., {Deming}, D., \& {Madhusudhan}, N. 2014,
  \apj, 795, 166, \dodoi{10.1088/0004-637X/795/2/166}

\bibitem[{{Cubillos} {et~al.}(2022){Cubillos}, {Harrington}, {Blecic}, {Himes},
  {Rojo}, {Loredo}, {Lust}, {Challener}, {Foster}, {Stemm}, {Foster}, \&
  {Blumenthal}}]{Cubillos2022}
{Cubillos}, P.~E., {Harrington}, J., {Blecic}, J., {et~al.} 2022, \psj, 3, 81,
  \dodoi{10.3847/PSJ/ac348b}

\bibitem[{{Dobbs-Dixon} \& {Blecic}(2022)}]{Dobbs-Dixon2022}
{Dobbs-Dixon}, I., \& {Blecic}, J. 2022, \apj, 929, 46,
  \dodoi{10.3847/1538-4357/ac5898}

\bibitem[{{Drummond} {et~al.}(2016){Drummond}, {Tremblin}, {Baraffe},
  {Amundsen}, {Mayne}, {Venot}, \& {Goyal}}]{Drummond2016}
{Drummond}, B., {Tremblin}, P., {Baraffe}, I., {et~al.} 2016, \aap, 594, A69,
  \dodoi{10.1051/0004-6361/201628799}

\bibitem[{Elsasser(1942)}]{Elsasser1942}
Elsasser, W. 1942, Heat Transfer by Infrared Radiation in the Atmosphere,
  Harvard meteorological studies (Harvard University, Blue Hill Meteorological
  Observatory).
\newblock \url{https://books.google.ch/books?id=u4p0vgAACAAJ}

\bibitem[{{Feng} {et~al.}(2020){Feng}, {Line}, \& {Fortney}}]{Feng2020}
{Feng}, Y.~K., {Line}, M.~R., \& {Fortney}, J.~J. 2020, \aj, 160, 137,
  \dodoi{10.3847/1538-3881/aba8f9}

\bibitem[{{Forget} \& {Pierrehumbert}(1997)}]{Forget1997}
{Forget}, F., \& {Pierrehumbert}, R.~T. 1997, Science, 278, 1273,
  \dodoi{10.1126/science.278.5341.1273}

\bibitem[{{Fu} \& {Liou}(1993)}]{Fu1993}
{Fu}, Q., \& {Liou}, K.~N. 1993, Journal of Atmospheric Sciences, 50, 2008,
  \dodoi{10.1175/1520-0469(1993)050<2008:POTRPO>2.0.CO;2}

\bibitem[{{Gandhi} \& {Madhusudhan}(2018)}]{Gandhi2018}
{Gandhi}, S., \& {Madhusudhan}, N. 2018, \mnras, 474, 271,
  \dodoi{10.1093/mnras/stx2748}

\bibitem[{{Gandhi} {et~al.}(2019){Gandhi}, {Madhusudhan}, {Hawker}, \&
  {Piette}}]{Gandhi2019}
{Gandhi}, S., {Madhusudhan}, N., {Hawker}, G., \& {Piette}, A. 2019, \aj, 158,
  228, \dodoi{10.3847/1538-3881/ab4efc}

\bibitem[{{Gao} {et~al.}(2018){Gao}, {Marley}, \& {Ackerman}}]{Gao2018}
{Gao}, P., {Marley}, M.~S., \& {Ackerman}, A.~S. 2018, \apj, 855, 86,
  \dodoi{10.3847/1538-4357/aab0a1}

\bibitem[{{Guillot}(2010)}]{Guillot2010}
{Guillot}, T. 2010, \aap, 520, A27, \dodoi{10.1051/0004-6361/200913396}

\bibitem[{He(1999)}]{He1999}
He, J.-H. 1999, International journal of non-linear mechanics, 34, 699

\bibitem[{{Heng} \& {Kitzmann}(2017)}]{Heng2017}
{Heng}, K., \& {Kitzmann}, D. 2017, \apjs, 232, 20,
  \dodoi{10.3847/1538-4365/aa8907}

\bibitem[{{Heng} {et~al.}(2014){Heng}, {Mendon{\c{c}}a}, \& {Lee}}]{Heng2014}
{Heng}, K., {Mendon{\c{c}}a}, J.~M., \& {Lee}, J.-M. 2014, \apjs, 215, 4,
  \dodoi{10.1088/0067-0049/215/1/4}

\bibitem[{{Heng} {et~al.}(2011){Heng}, {Menou}, \& {Phillipps}}]{Heng2011}
{Heng}, K., {Menou}, K., \& {Phillipps}, P.~J. 2011, \mnras, 413, 2380,
  \dodoi{10.1111/j.1365-2966.2011.18315.x}

\bibitem[{{Henyey} \& {Greenstein}(1941)}]{Henyey1941}
{Henyey}, L.~G., \& {Greenstein}, J.~L. 1941, \apj, 93, 70,
  \dodoi{10.1086/144246}

\bibitem[{{Iro} {et~al.}(2005){Iro}, {B{\'e}zard}, \& {Guillot}}]{Iro2005}
{Iro}, N., {B{\'e}zard}, B., \& {Guillot}, T. 2005, \aap, 436, 719,
  \dodoi{10.1051/0004-6361:20048344}

\bibitem[{{Irwin} {et~al.}(2020){Irwin}, {Parmentier}, {Taylor}, {Barstow},
  {Aigrain}, {Lee}, \& {Garland}}]{Irwin2020}
{Irwin}, P. G.~J., {Parmentier}, V., {Taylor}, J., {et~al.} 2020, \mnras, 493,
  106, \dodoi{10.1093/mnras/staa238}

\bibitem[{{Kataria} {et~al.}(2013){Kataria}, {Showman}, {Lewis}, {Fortney},
  {Marley}, \& {Freedman}}]{Kataria2013}
{Kataria}, T., {Showman}, A.~P., {Lewis}, N.~K., {et~al.} 2013, \apj, 767, 76,
  \dodoi{10.1088/0004-637X/767/1/76}

\bibitem[{{Kitzmann}(2016)}]{Kitzmann2016}
{Kitzmann}, D. 2016, \apjl, 817, L18, \dodoi{10.3847/2041-8205/817/2/L18}

\bibitem[{{Kitzmann} \& {Heng}(2018)}]{Kitzmann2018}
{Kitzmann}, D., \& {Heng}, K. 2018, \mnras, 475, 94,
  \dodoi{10.1093/mnras/stx3141}

\bibitem[{{Kitzmann} {et~al.}(2020){Kitzmann}, {Heng}, {Oreshenko}, {Grimm},
  {Apai}, {Bowler}, {Burgasser}, \& {Marley}}]{Kitzmann2020}
{Kitzmann}, D., {Heng}, K., {Oreshenko}, M., {et~al.} 2020, \apj, 890, 174,
  \dodoi{10.3847/1538-4357/ab6d71}

\bibitem[{{Kitzmann} {et~al.}(2013){Kitzmann}, {Patzer}, \&
  {Rauer}}]{Kitzmann2013}
{Kitzmann}, D., {Patzer}, A.~B.~C., \& {Rauer}, H. 2013, \aap, 557, A6,
  \dodoi{10.1051/0004-6361/201220025}

\bibitem[{{Komacek} {et~al.}(2022){Komacek}, {Tan}, {Gao}, \&
  {Lee}}]{Komacek2022}
{Komacek}, T.~D., {Tan}, X., {Gao}, P., \& {Lee}, E. K.~H. 2022, \apj, 934, 79,
  \dodoi{10.3847/1538-4357/ac7723}

\bibitem[{{Kreidberg} {et~al.}(2014){Kreidberg}, {Bean}, {D{\'e}sert}, {Line},
  {Fortney}, {Madhusudhan}, {Stevenson}, {Showman}, {Charbonneau},
  {McCullough}, {Seager}, {Burrows}, {Henry}, {Williamson}, {Kataria}, \&
  {Homeier}}]{Kreidberg2014}
{Kreidberg}, L., {Bean}, J.~L., {D{\'e}sert}, J.-M., {et~al.} 2014, \apjl, 793,
  L27, \dodoi{10.1088/2041-8205/793/2/L27}

\bibitem[{{Kylling} {et~al.}(1995){Kylling}, {Stamnes}, \&
  {Tsay}}]{Kylling1995}
{Kylling}, A., {Stamnes}, K., \& {Tsay}, S.~C. 1995, Journal of Atmospheric
  Chemistry, 21, 115, \dodoi{10.1007/BF00696577}

\bibitem[{{Lee}(2023)}]{Lee2023}
{Lee}, E. K.~H. 2023, \mnras, 524, 2918, \dodoi{10.1093/mnras/stad2037}

\bibitem[{{Lee} {et~al.}(2021){Lee}, {Parmentier}, {Hammond}, {Grimm},
  {Kitzmann}, {Tan}, {Tsai}, \& {Pierrehumbert}}]{Lee2021}
{Lee}, E. K.~H., {Parmentier}, V., {Hammond}, M., {et~al.} 2021, \mnras, 506,
  2695, \dodoi{10.1093/mnras/stab1851}

\bibitem[{{Li}(2002)}]{Li2002}
{Li}, J. 2002, Journal of Atmospheric Sciences, 59, 3302,
  \dodoi{10.1175/1520-0469(2002)059<3302:AFUCIA>2.0.CO;2}

\bibitem[{{Li} \& {Fu}(2000)}]{Li2000}
{Li}, J., \& {Fu}, Q. 2000, Journal of Atmospheric Sciences, 57, 2905,
  \dodoi{10.1175/1520-0469(2000)057<2905:AAWSEF>2.0.CO;2}

\bibitem[{{Lin} {et~al.}(2018){Lin}, {Chen}, {Fan}, {Li}, {Stamnes}, \&
  {Stamnes}}]{Lin2018}
{Lin}, Z., {Chen}, N., {Fan}, Y., {et~al.} 2018, Journal of the Atmospheric
  Sciences, 75, 327, \dodoi{10.1175/JAS-D-17-0233.1}

\bibitem[{{Line} {et~al.}(2014){Line}, {Knutson}, {Wolf}, \& {Yung}}]{Line2014}
{Line}, M.~R., {Knutson}, H., {Wolf}, A.~S., \& {Yung}, Y.~L. 2014, \apj, 783,
  70, \dodoi{10.1088/0004-637X/783/2/70}

\bibitem[{{Line} {et~al.}(2013){Line}, {Wolf}, {Zhang}, {Knutson}, {Kammer},
  {Ellison}, {Deroo}, {Crisp}, \& {Yung}}]{Line2013}
{Line}, M.~R., {Wolf}, A.~S., {Zhang}, X., {et~al.} 2013, \apj, 775, 137,
  \dodoi{10.1088/0004-637X/775/2/137}

\bibitem[{{Malik} {et~al.}(2019){Malik}, {Kitzmann}, {Mendon{\c{c}}a}, {Grimm},
  {Marleau}, {Linder}, {Tsai}, \& {Heng}}]{Malik2019}
{Malik}, M., {Kitzmann}, D., {Mendon{\c{c}}a}, J.~M., {et~al.} 2019, \aj, 157,
  170, \dodoi{10.3847/1538-3881/ab1084}

\bibitem[{{Marley} {et~al.}(2021){Marley}, {Saumon}, {Visscher}, {Lupu},
  {Freedman}, {Morley}, {Fortney}, {Seay}, {Smith}, {Teal}, \&
  {Wang}}]{Marley2021}
{Marley}, M.~S., {Saumon}, D., {Visscher}, C., {et~al.} 2021, \apj, 920, 85,
  \dodoi{10.3847/1538-4357/ac141d}

\bibitem[{{Mikal-Evans} {et~al.}(2022){Mikal-Evans}, {Sing}, {Barstow},
  {Kataria}, {Goyal}, {Lewis}, {Taylor}, {Mayne}, {Daylan}, {Wakeford},
  {Marley}, \& {Spake}}]{Mikal-Evans2022}
{Mikal-Evans}, T., {Sing}, D.~K., {Barstow}, J.~K., {et~al.} 2022, Nature
  Astronomy, 6, 471, \dodoi{10.1038/s41550-021-01592-w}

\bibitem[{{Molli{\`e}re} {et~al.}(2017){Molli{\`e}re}, {van Boekel}, {Bouwman},
  {Henning}, {Lagage}, \& {Min}}]{Molliere2017}
{Molli{\`e}re}, P., {van Boekel}, R., {Bouwman}, J., {et~al.} 2017, \aap, 600,
  A10, \dodoi{10.1051/0004-6361/201629800}

\bibitem[{{Parmentier} {et~al.}(2015){Parmentier}, {Guillot}, {Fortney}, \&
  {Marley}}]{Parmentier2015}
{Parmentier}, V., {Guillot}, T., {Fortney}, J.~J., \& {Marley}, M.~S. 2015,
  \aap, 574, A35, \dodoi{10.1051/0004-6361/201323127}

\bibitem[{{Pierrehumbert}(2010)}]{Pierrehumbert2010}
{Pierrehumbert}, R.~T. 2010, {Principles of Planetary Climate}

\bibitem[{{Powell} {et~al.}(2018){Powell}, {Zhang}, {Gao}, \&
  {Parmentier}}]{Powell2018}
{Powell}, D., {Zhang}, X., {Gao}, P., \& {Parmentier}, V. 2018, \apj, 860, 18,
  \dodoi{10.3847/1538-4357/aac215}

\bibitem[{{Roman} {et~al.}(2021){Roman}, {Kempton}, {Rauscher}, {Harada},
  {Bean}, \& {Stevenson}}]{Roman2021}
{Roman}, M.~T., {Kempton}, E. M.~R., {Rauscher}, E., {et~al.} 2021, \apj, 908,
  101, \dodoi{10.3847/1538-4357/abd549}

\bibitem[{{Rooney} {et~al.}(2024){Rooney}, {Batalha}, \& {Marley}}]{Rooney2024}
{Rooney}, C.~M., {Batalha}, N.~E., \& {Marley}, M.~S. 2024, \apj, 960, 131,
  \dodoi{10.3847/1538-4357/ad05c5}

\bibitem[{{Schneider} {et~al.}(2022){Schneider}, {Carone}, {Decin},
  {J{\o}rgensen}, {Molli{\`e}re}, {Baeyens}, {Kiefer}, \&
  {Helling}}]{Schneider2022}
{Schneider}, A.~D., {Carone}, L., {Decin}, L., {et~al.} 2022, \aap, 664, A56,
  \dodoi{10.1051/0004-6361/202142728}

\bibitem[{{Showman} {et~al.}(2009){Showman}, {Fortney}, {Lian}, {Marley},
  {Freedman}, {Knutson}, \& {Charbonneau}}]{Showman2009}
{Showman}, A.~P., {Fortney}, J.~J., {Lian}, Y., {et~al.} 2009, \apj, 699, 564,
  \dodoi{10.1088/0004-637X/699/1/564}

\bibitem[{{Sing} {et~al.}(2016){Sing}, {Fortney}, {Nikolov}, {Wakeford},
  {Kataria}, {Evans}, {Aigrain}, {Ballester}, {Burrows}, {Deming},
  {D{\'e}sert}, {Gibson}, {Henry}, {Huitson}, {Knutson}, {Lecavelier Des
  Etangs}, {Pont}, {Showman}, {Vidal-Madjar}, {Williamson}, \&
  {Wilson}}]{Sing2016}
{Sing}, D.~K., {Fortney}, J.~J., {Nikolov}, N., {et~al.} 2016, \nat, 529, 59,
  \dodoi{10.1038/nature16068}

\bibitem[{{Stamnes} {et~al.}(1988){Stamnes}, {Tsay}, {Jayaweera}, \&
  {Wiscombe}}]{Stamnes1988}
{Stamnes}, K., {Tsay}, S.~C., {Jayaweera}, K., \& {Wiscombe}, W. 1988, \ao, 27,
  2502, \dodoi{10.1364/AO.27.002502}

\bibitem[{Stamnes {et~al.}(2000)Stamnes, Tsay, Wiscombe, \&
  Laszlo}]{Stamnes2000}
Stamnes, K., Tsay, S.-C., Wiscombe, W., \& Laszlo, I. 2000

\bibitem[{{Stevenson} {et~al.}(2014){Stevenson}, {D{\'e}sert}, {Line}, {Bean},
  {Fortney}, {Showman}, {Kataria}, {Kreidberg}, {McCullough}, {Henry},
  {Charbonneau}, {Burrows}, {Seager}, {Madhusudhan}, {Williamson}, \&
  {Homeier}}]{Stevenson2014}
{Stevenson}, K.~B., {D{\'e}sert}, J.-M., {Line}, M.~R., {et~al.} 2014, Science,
  346, 838, \dodoi{10.1126/science.1256758}

\bibitem[{{Tan} \& {Showman}(2021)}]{Tan2021}
{Tan}, X., \& {Showman}, A.~P. 2021, \mnras, 502, 678,
  \dodoi{10.1093/mnras/stab060}

\bibitem[{{Taylor} \& {Parmentier}(2023)}]{Taylor2023}
{Taylor}, J., \& {Parmentier}, V. 2023, \mnras, \dodoi{10.1093/mnras/stad2287}

\bibitem[{{Taylor} {et~al.}(2021){Taylor}, {Parmentier}, {Line}, {Lee},
  {Irwin}, \& {Aigrain}}]{Taylor2021}
{Taylor}, J., {Parmentier}, V., {Line}, M.~R., {et~al.} 2021, \mnras, 506,
  1309, \dodoi{10.1093/mnras/stab1854}

\bibitem[{{Toon} {et~al.}(1989){Toon}, {McKay}, {Ackerman}, \&
  {Santhanam}}]{Toon1989}
{Toon}, O.~B., {McKay}, C.~P., {Ackerman}, T.~P., \& {Santhanam}, K. 1989,
  \jgr, 94, 16287, \dodoi{10.1029/JD094iD13p16287}

\bibitem[{{Waldmann} {et~al.}(2015){Waldmann}, {Rocchetto}, {Tinetti},
  {Barton}, {Yurchenko}, \& {Tennyson}}]{Waldmann2015}
{Waldmann}, I.~P., {Rocchetto}, M., {Tinetti}, G., {et~al.} 2015, \apj, 813,
  13, \dodoi{10.1088/0004-637X/813/1/13}

\bibitem[{{Wiscombe}(1977)}]{Wiscombe1977}
{Wiscombe}, W.~J. 1977, Journal of Atmospheric Sciences, 34, 1408,
  \dodoi{10.1175/1520-0469(1977)034<1408:TDMRYA>2.0.CO;2}

\bibitem[{{Wolf} {et~al.}(2022){Wolf}, {Kopparapu}, {Haqq-Misra}, \&
  {Fauchez}}]{Wolf2022}
{Wolf}, E.~T., {Kopparapu}, R., {Haqq-Misra}, J., \& {Fauchez}, T.~J. 2022,
  \psj, 3, 7, \dodoi{10.3847/PSJ/ac3f3d}

\bibitem[{{Zhang} {et~al.}(2017){Zhang}, {Shi}, {Li}, {Wu}, \&
  {Iwabuchi}}]{Zhang2017}
{Zhang}, F., {Shi}, Y.-N., {Li}, J., {Wu}, K., \& {Iwabuchi}, H. 2017, Journal
  of Atmospheric Sciences, 74, 419, \dodoi{10.1175/JAS-D-16-0172.1}

\end{thebibliography}
\bibliographystyle{aasjournal}

\clearpage
\appendix

\section{AA vs EAA}
\label{app:AA_EAA}

In Figure \ref{fig:AA_hrate} we show the errors between the AA and EAA methods in vertical heating rates compared to the 64 stream DISORT reference values.
These results generally show an improvement of EAA over AA, in particular for the deep cloud scenario.
This also shows the superiority of the four-stream variants over the two-stream, the four-steam variants show a much increased accuracy over the two-steam, in particular at the low pressure regions.
We can therefore recommend using the four-stream EAA variant for modelling calculations and as the zeroth order estimate for the VIM method.

\begin{figure*}
    \centering
    \includegraphics[width=0.49\textwidth]{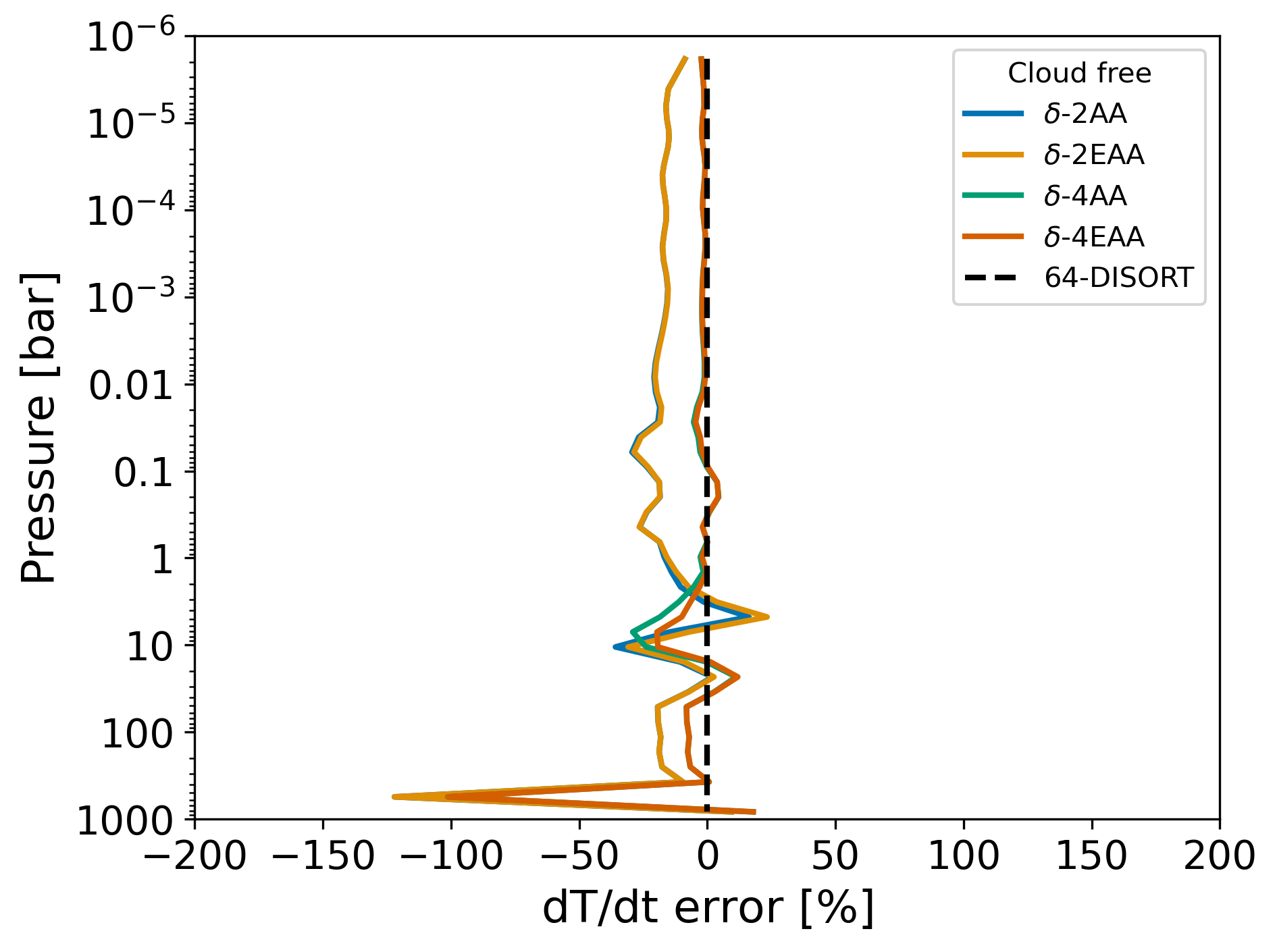}
    \includegraphics[width=0.49\textwidth]{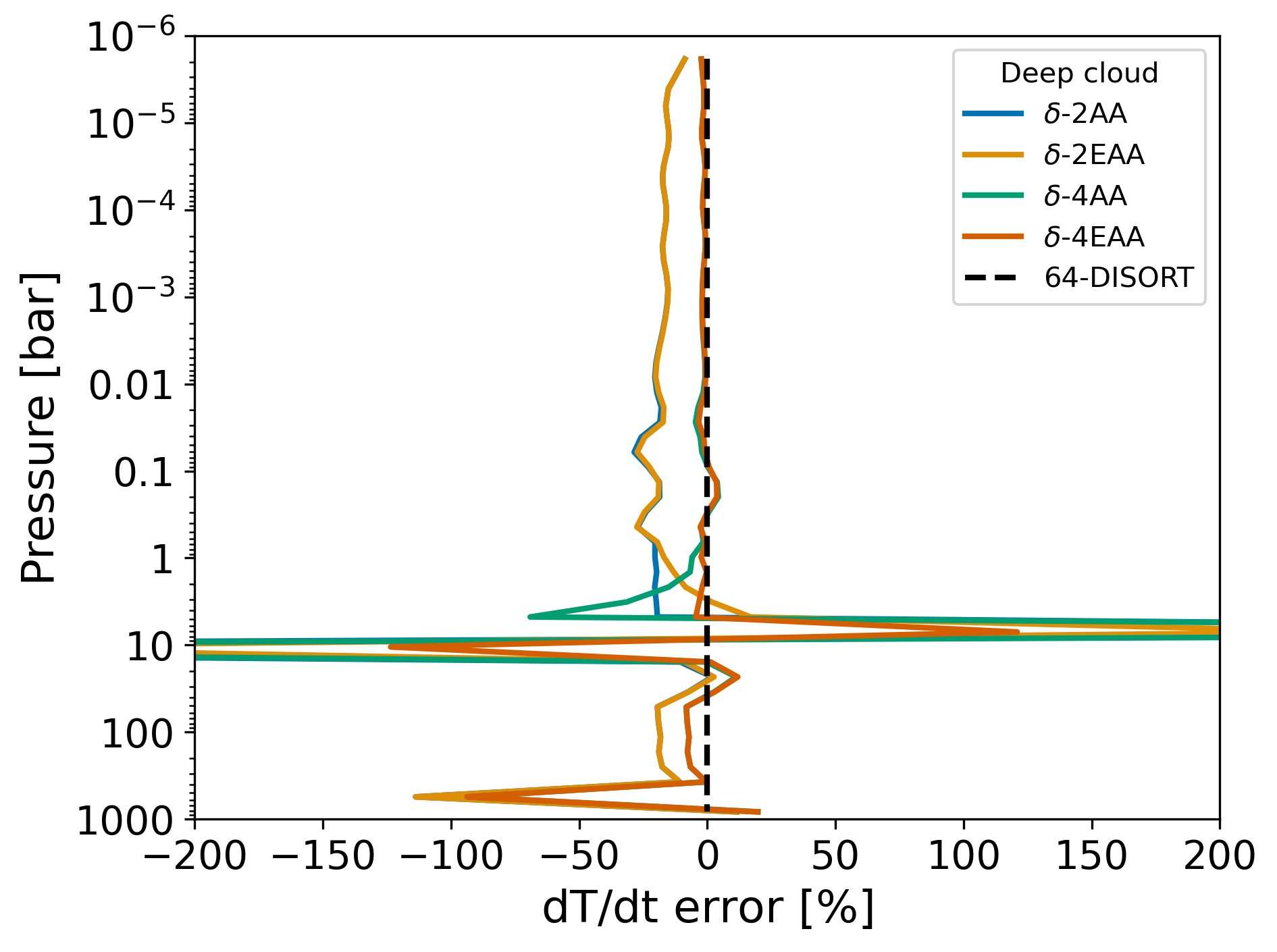}
     \includegraphics[width=0.49\textwidth]{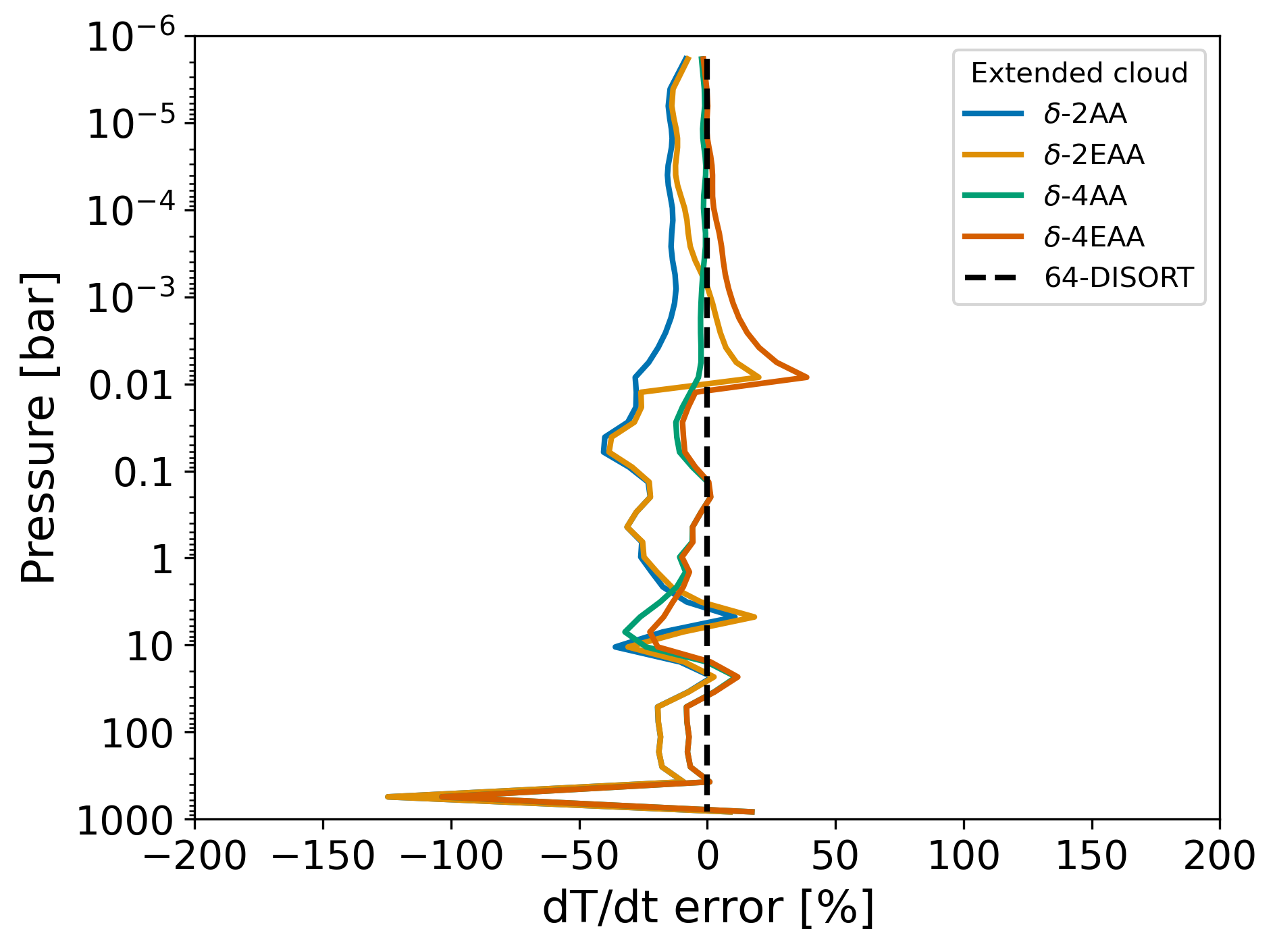}
    \includegraphics[width=0.49\textwidth]{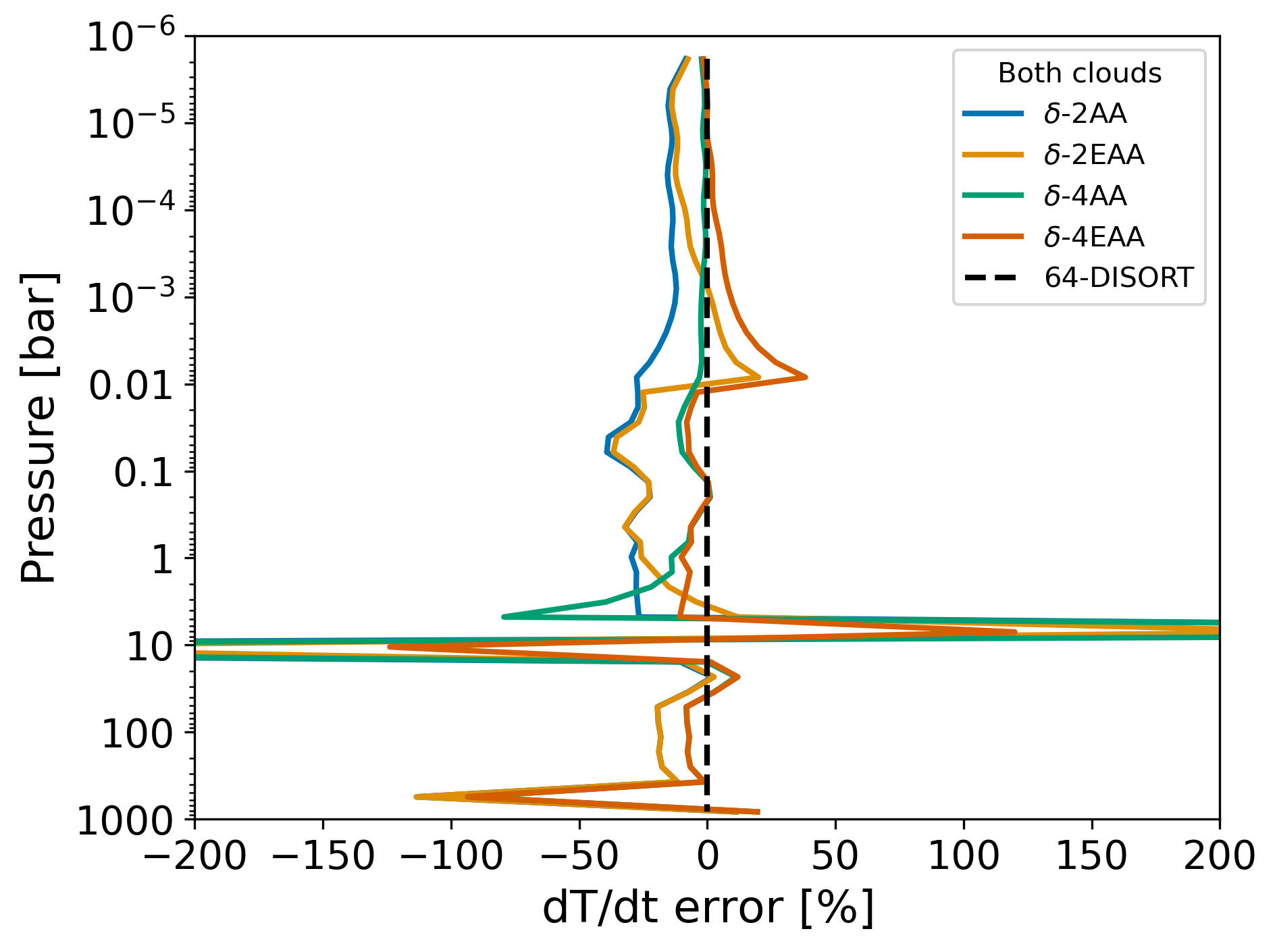}
    \caption{Percent error in the heating rates for the 2 and 4 stream AA and EAA methods.
    Each method is shown as a coloured solid line.}
    \label{fig:AA_hrate}
\end{figure*}

\end{document}